\documentclass[aps,prl,
superscriptaddress,
showpacs,
preprintnumbers,
bibnotes,
amsmath,
amssymb,
twocolumn,
floatfix,
]{revtex4-1}
\usepackage[utf8]{inputenc}
\usepackage{array}
\usepackage[normalem]{ulem}
\newcolumntype{L}[1]{>{\raggedright\let\newline\\\arraybackslash\hspace{0pt}}m{#1}}
\newcolumntype{C}[1]{>{\centering\let\newline\\\arraybackslash\hspace{0pt}}m{#1}}
\newcolumntype{R}[1]{>{\raggedleft\let\newline\\\arraybackslash\hspace{0pt}}m{#1}}

\newcommand{\EQ}{\overline{\rm Q}}
\newcommand{\SQ}{\overline{\rm S}}

\usepackage[caption=false]{subfig}

\usepackage{graphicx}
\usepackage{dcolumn}
\usepackage{bm}
\usepackage[usenames,dvipsnames]{color}
\usepackage{verbatim}
\usepackage{amsfonts}
\usepackage{longtable}
\usepackage{natbib}
\usepackage{hyperref}
\usepackage{dcolumn}

\newcolumntype{d}[1]{D{;}{.}{#1}}

\graphicspath{{figs/}}
\def\slashed#1{\kern+0.10em /\kern-0.50em #1}

\begin{document}
\title{Calculation of the hadronic vacuum polarization contribution to the muon anomalous magnetic moment}

\newcommand\bnl{Physics Department, Brookhaven National Laboratory, Upton, NY 11973, USA}
\newcommand\cu{Physics Department, Columbia University, New York, NY 10027, USA}
\newcommand\pu{School of Computing \& Mathematics, Plymouth University, Plymouth PL4 8AA, UK}
\newcommand\riken{RIKEN-BNL Research Center, Brookhaven National Laboratory, Upton, NY 11973, USA}
\newcommand\edinb{School of Physics and Astronomy, The University of Edinburgh, Edinburgh EH9 3FD, UK}
\newcommand\uconn{Physics Department, University of Connecticut, Storrs, CT 06269-3046, USA}
\newcommand\soton{School of Physics and Astronomy, University of Southampton,  Southampton SO17 1BJ, UK}
\newcommand\york{Mathematics \& Statistics, York University, Toronto, ON, M3J 1P3, Canada}
\newcommand\cssm{CSSM, University of Adelaide, Adelaide 5005 SA, Australia}
\newcommand\cern{CERN, Physics Department, 1211 Geneva 23, Switzerland}

\author{T.~Blum}\affiliation{\uconn}
\author{P.A.~Boyle}\affiliation{\edinb}
\author{V.~G\"ulpers}\affiliation{\soton}
\author{T.~Izubuchi}\affiliation{\bnl}\affiliation{\riken}
\author{L.~Jin}\affiliation{\uconn}\affiliation{\riken}
\author{C.~Jung}\affiliation{\bnl}
\author{A.~J\"uttner}\affiliation{\soton}
\author{C.~Lehner}\thanks{Corresponding author}\email{clehner@quark.phy.bnl.gov}\affiliation{\bnl}
\author{A.~Portelli}\affiliation{\edinb}
\author{J.T.~Tsang}\affiliation{\edinb}

\collaboration{RBC and UKQCD Collaborations}
\noaffiliation

\date{January 22, 2018}

\begin{abstract} 
  We present a first-principles lattice QCD+QED calculation at
  physical pion mass of the leading-order hadronic vacuum polarization contribution
  to the muon anomalous magnetic moment.  The total contribution of
  up, down, strange, and charm quarks including QED and strong isospin
  breaking effects is found to be $a_\mu^{\rm HVP~LO}
  =715.4(16.3)(9.2) \times 10^{-10}$, where the first error is
  statistical and the second is systematic.  By supplementing lattice
  data for very short and long distances with experimental R-ratio
  data using the compilation of Ref.~\cite{Jegerlehner2017}, we
  significantly improve the precision of our calculation and find
  $a_\mu^{\rm HVP~LO} = 692.5(1.4)(0.5)(0.7)(2.1) \times 10^{-10}$
  with lattice statistical, lattice systematic, R-ratio statistical,
  and R-ratio systematic errors given separately.  This is the
  currently most precise determination of the leading-order hadronic
  vacuum polarization contribution to the muon anomalous magnetic
  moment.  In addition, we present the first lattice calculation of
  the light-quark QED correction at physical pion mass.
\end{abstract}

\pacs{
      12.38.Gc  
}

\preprint{}

\keywords{anomalous magnetic moment, muon, R-ratio, lattice QCD} 
\maketitle

\section{Introduction}
The anomalous magnetic moment of the muon $a_\mu$ is defined as the
deviation of the Land\'e factor $g_\mu$ from Dirac's relativistic
quantum mechanics result, $a_\mu = \frac{g_\mu -2}{2}$.
It is one of the most precisely determined quantities in particle
physics and is currently known both experimentally (BNL E821)
\cite{Bennett:2006fi} and from a standard model theory calculation \cite{PDG2017} to approximately
$1/2$ parts per million.

Interestingly, the standard model result $a_\mu^{\rm SM}$
deviates from the experimental measurement $a_\mu^{\rm EXP}$ 
at the 3--4 sigma level, depending on which determination of the leading-order
hadronic vacuum polarization $a_\mu^{\rm HVP~LO}$ is used.  One finds
\begin{align}\label{eqn:tension}
  a_\mu^{\rm EXP} - a_\mu^{\rm SM} = \,\,& 25.0(4.3)(2.6)(6.3) \times 10 ^ {-10}~\text{\cite{Hagiwara:2011af}} \,,\notag\\
 & 31.8 (4.1)(2.6)(6.3)\times 10 ^ {-10}~\text{\cite{Jegerlehner:2017lbd}} \,,\notag\\
  & 26.8 (3.4)(2.6)(6.3) \times 10 ^ {-10}~\text{\cite{Davier:2017zfy}} \,,
\end{align}
where the quoted errors correspond to the uncertainty in $a_\mu^{\rm
  HVP~LO}$, $a_\mu^{\rm SM} - a_\mu^{\rm HVP~LO}$, and $a_\mu^{\rm
  EXP}$.  This tension may hint at new physics beyond the standard
model of particle physics such that a reduction of uncertainties in
Eq.~\eqref{eqn:tension} is highly desirable.  New experiments at
Fermilab (E989) \cite{Carey:2009zzb} and J-PARC (E34)
\cite{Aoki:2009xxx} intend to decrease the experimental uncertainty by
a factor of four.  First results of the E989 experiment may be available 
before the end of 2018 \cite{PrivateComm} such that a
reduction in uncertainty of the $a_\mu^{\rm HVP~LO}$ contribution is
of timely interest.

In the following, we perform a complete first-principles calculation of $a_\mu^{\rm
  HVP~LO}$ in lattice QCD+QED at physical pion mass with
non-degenerate up and down quark masses and present results for the
up, down, strange, and charm quark contributions.  Our lattice calculation of
the light-quark QED correction to $a_\mu^{\rm HVP~LO}$ is the
first such calculation performed at physical pion mass.  In addition, we
replace lattice data at very short and long distances by experimental
$e^+e^-$ scattering data using the compilation of
Ref.~\cite{Jegerlehner2017}, which allows us to produce the currently
most precise determination of $a_\mu^{\rm HVP~LO}$.

\section{Computational Method}
The general setup of our non-perturbative lattice computation is
described in Ref.~\cite{Blum:2002ii}.  We compute
\begin{align}\label{eqn:defamu}
  a_\mu = 4 \alpha^2 \int_0^\infty dq^2 f(q^2) [ \Pi(q^2) - \Pi(q^2=0) ] \,,
\end{align}
where $f(q^2)$ is a known analytic function \cite{Blum:2002ii} and
$\Pi(q^2)$ is defined as $ \sum_x e^{i q x} \langle J_\mu(x) J_\nu(0)
\rangle = (\delta_{\mu\nu} q^2 - q_\mu q_\nu) \Pi(q^2) $ with sum over
space-time coordinate $x$ and $ J_\mu(x) = i\sum_f Q_f
\overline{\Psi}_f(x) \gamma_\mu \Psi_f(x)$.  The sum is over up, down,
strange, and charm quark flavors with QED charges $Q_{\rm
  up,~charm}=2/3$ and $Q_{\rm down,~strange}=-1/3$.  For convenience
we do not explicitly write the superscript HVP~LO.  We compute
$\Pi(q^2)$ using the kernel function of
Refs.~\cite{Bernecker:2011gh,Feng:2013xsa}
\begin{align}\label{eqn:wick}
  \Pi(q^2) - \Pi(q^2=0) &= \sum_{t} \left( \frac{\cos(q t) - 1}{q^2} + \frac12 t^2\right) C(t)
\end{align}
with
$  C(t) = \frac13 \sum_{\vec{x}}\sum_{j=0,1,2} \langle J_j(\vec{x},t) J_j(0) \rangle $.
With appropriate definition of $w_t$, we can therefore write
\begin{align}
  a_\mu = \sum_t w_t C(t) \,.
\end{align}

The correlator $C(t)$ is computed in lattice QCD+QED with dynamical
up, down, and strange quarks and non-degenerate up and down quark
masses.  We compute the missing contributions to $a_\mu$ from bottom quarks and
from charm sea quarks in perturbative QCD
\cite{Harlander:2002ur} by integrating the time-like region above $2$ GeV
and find them to be smaller than $0.3 \times 10^{-10}$.

We tune the bare up, down, and strange quark masses $m_{\rm up}$, $m_{\rm
  down}$, and $m_{\rm strange}$ such that the $\pi^0$, $\pi^+$, $K^0$,
and $K^+$ meson masses computed in our calculation agree with the
respective experimental measurements \footnote{We minimize the sum of
  squared differences of computed and measured meson masses.}.  The
lattice spacing is determined by setting the $\Omega^-$ mass to its
experimental value.  We perform the calculation as a perturbation
around an isospin-symmetric lattice QCD computation
\cite{deDivitiis:2011eh,deDivitiis:2013xla} with two degenerate light
quarks with mass $m_{\rm light}$ and a heavy quark with mass $m_{\rm
  heavy}$ tuned to produce a pion mass of 135.0 MeV and a kaon mass of
495.7 MeV \cite{Blum:2014tka}.  The correlator is expanded in the
fine-structure constant $\alpha$ as well as $\Delta m_{\rm up,~down} =
m_{\rm up,~down} - m_{\rm light}$, and $\Delta m_{\rm strange} =
m_{\rm strange} - m_{\rm heavy}$.  We write
\begin{align}
  C(t) &= C^{(0)}(t) + \alpha C^{(1)}_{\rm QED}(t)+ \sum_{f}\Delta m_f C^{(1)}_{\rm \Delta m_f}(t) \notag\\
 &\quad + {\cal O}(\alpha^2,\alpha \Delta m, \Delta m^2) \,,
\end{align}
where $C^{(0)}(t)$ is obtained in the lattice QCD calculation
at the isospin symmetric point and the expansion terms define the
QED and strong isospin-breaking (SIB) corrections, respectively.  We keep
only the leading corrections in $\alpha$ and $\Delta m_f$ which is
sufficient for the desired precision.

We insert the photon-quark vertices perturbatively with photons
coupled to local lattice vector currents multiplied by the
renormalization factor $Z_V$ \cite{Blum:2014tka}.  We use $Z_A\approx
Z_V$ for the charm \cite{Boyle:2017jwu} and QED corrections.  The SIB
correction is computed by inserting scalar operators in the respective
quark lines.  The procedure used for effective masses in such a
perturbative expansion is explained in Ref.~\cite{Boyle:2017gzv}.  We
use the finite-volume QED$_{\rm L}$ prescription
\cite{Hayakawa:2008an} and remove the universal $1/L$ and $1/L^2$
corrections to the masses \cite{Borsanyi:2014jba} with spatial lattice
size $L$.  The effect of $1/L^3$ corrections is small compared to our
statistical uncertainties.  We find
 $ \Delta m_{\rm up} = -0.00050(1)$, $\Delta m_{\rm down} = 0.00050(1)$, and $\Delta m_{\rm strange}= -0.0002(2)$
for the 48I lattice ensemble described in Ref.~\cite{Blum:2014tka}.
The shift of the $\Omega^-$ mass due to the QED correction is
significantly smaller than the lattice spacing uncertainty and its
effect on $C(t)$ is therefore not included separately.

\begin{figure}[tbh] 
  \centering
  \includegraphics[height=2cm]{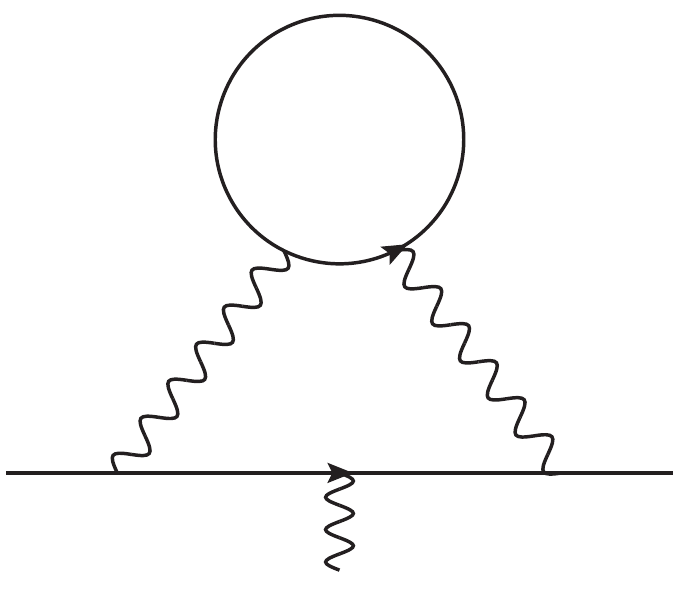}
  \hspace{1cm}
 \includegraphics[height=2cm]{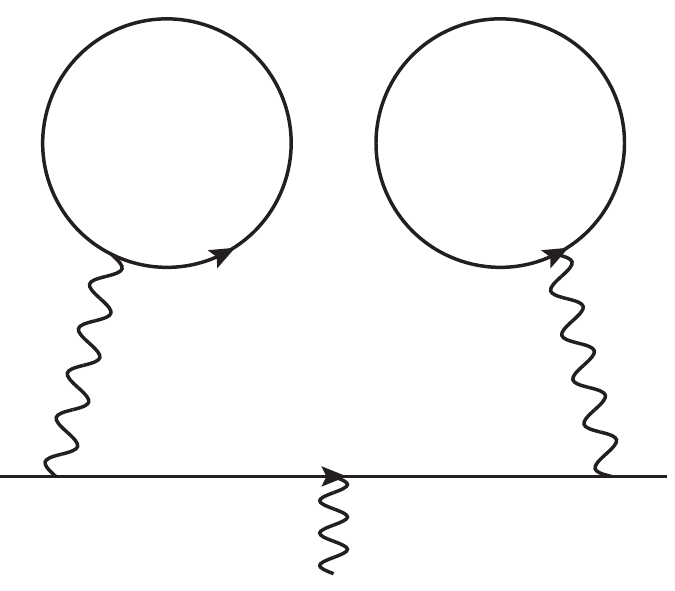}
 \caption{Quark-connected (left) and quark-disconnected (right)
   diagram for the calculation of $a_\mu^{\rm HVP~LO}$. We do not draw
   gluons but consider each diagram to represent all orders in QCD.}
  \label{fig:diag0}
\end{figure}

\begin{figure}[bth] 
  \centering
  \includegraphics[width=8.5cm]{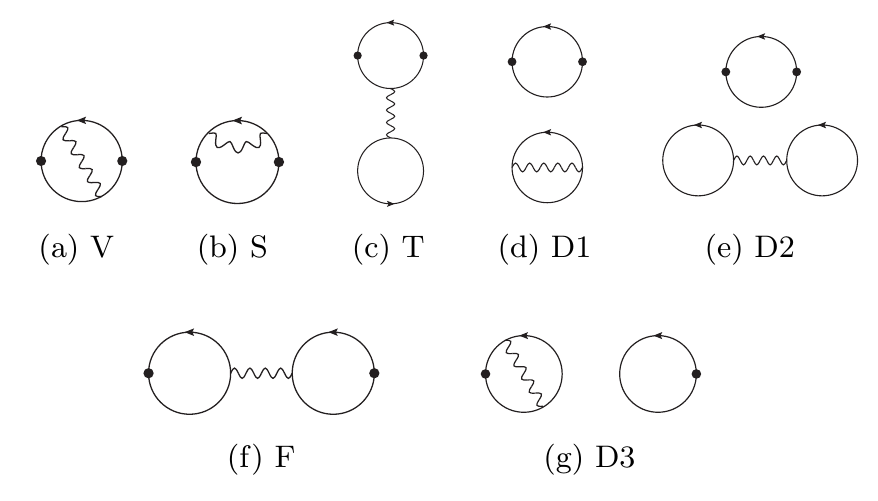}
  \caption{QED-correction diagrams with external pseudo-scalar or
    vector operators.}
  \label{fig:diagqed}
\end{figure}

Figure~\ref{fig:diag0} shows the quark-connected and
quark-disconnected contributions to $C^{(0)}$.  Similarly,
Fig.~\ref{fig:diagqed} shows the relevant diagrams for the QED
correction to the meson spectrum and the hadronic vacuum polarization.
The external vertices are pseudo-scalar operators for the former and
vector operators for the latter.  We refer to diagrams S and V as the
QED-connected and to diagram F as the QED-disconnected contribution.
We note that only the parts of diagram F with additional gluons
exchanged between the two quark loops contribute to $a_\mu^{\rm
  HVP~LO}$ as otherwise an internal cut through a single photon line
is possible.  For this reason, we subtract the separate
quantum-averages of quark loops in diagram F.  In the current
calculation, we neglect diagrams T, D1, D2, and D3. This approximation is estimated to yield
an ${\cal O}(10\%)$ correction for isospin splittings
\cite{Borsanyi:2013lga} for which the neglected diagrams are both
SU(3) and $1/N_c$ suppressed.  For the hadronic vacuum polarization
the contribution of neglected diagrams is still $1/N_c$ suppressed and
we adopt a corresponding $30\%$ uncertainty.

\begin{figure}[bt] 
  \centering
  \includegraphics[width=6cm]{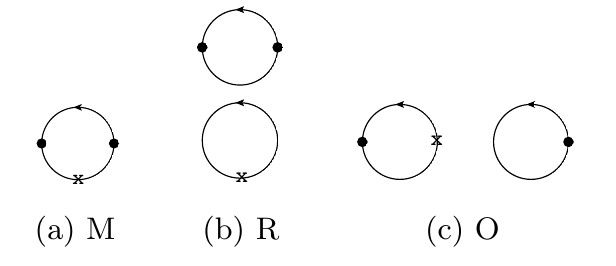}
  \caption{Strong isospin-breaking correction diagrams.  The crosses
    denote the insertion of a scalar operator.}
  \label{fig:diagsib}
\end{figure}

In Fig.~\ref{fig:diagsib}, we show the SIB
diagrams.  In the calculation presented here, we only include diagram
M.  For the meson masses this corresponds to neglecting the sea
quark mass correction, which we have previously \cite{Blum:2014tka} determined
to be an ${\cal O}(2\%)$ and ${\cal O}(14\%)$ effect for the pions and kaons, respectively.
This estimate is based on the analytic fits of (H7) and (H9) of Ref.~\cite{Blum:2014tka}
with ratios $C_2^{m_{\pi,~K}} / C_1^{m_{\pi,~K}}$ given in Tab.~XVII of the same reference.  For the hadronic vacuum polarization the
contribution of diagram R is negligible since $\Delta m_{\rm up}
\approx -\Delta m_{\rm down}$ and diagram O is SU(3) and $1/N_c$
suppressed.  We therefore assign a corresponding $10\%$
uncertainty to the SIB correction.

%
%
%

We also compute the ${\cal O}(\alpha)$ correction to the vector current
renormalization factor $Z_V$ used in $C^{(0)}$
\cite{Blum:2014tka,Boyle:2017gzv} and find a small correction of
approximately $0.05\%$ for the light quarks.

We perform the calculation of $C^{(0)}$ on the 48I and 64I ensembles
described in Ref.~\cite{Blum:2014tka} for the up, down, and strange
quark-connected contributions.  For the charm contribution we also
perform a global fit using additional ensembles described in
Ref.~\cite{Boyle:2017jwu}.  The quark-disconnected contribution as
well as QED and SIB corrections are computed only on ensemble 48I.

For the noisy light quark connected contribution, we employ a
multi-step approximation scheme with low-mode averaging
\cite{DeGrand:2004wh} over the entire volume and two levels of
approximations in a truncated deflated solver (AMA)
\cite{Collins:2007mh,Bali:2009hu,Blum:2012uh,Shintani:2014vja} of
randomly positioned point sources.  The low-mode space is generated
using a new Lanczos method working on multiple grids
\cite{Clark:2017wom}.  Our improved statistical estimator for the
quark disconnected diagrams is described in
Ref.~\cite{Blum:2015you} and our strategy for the strange quark is
published in Ref.~\cite{Blum:2016xpd}.  For diagram F, we re-use
point-source propagators generated in Ref.~\cite{Blum:2016lnc}.

The correlator $C(t)$ is related to the R-ratio data \cite{Bernecker:2011gh} by
 $ C(t) =\frac{1}{12 \pi^2} \int_0^\infty d(\sqrt{s}) R(s) s e^{-\sqrt{s} t} $
with $R(s) =\frac{3 s}{4 \pi \alpha^2} \sigma(s,e^+e^- \to {\rm had})$.
\begin{figure}[tb] 
  \centering
\includegraphics[scale=0.55]{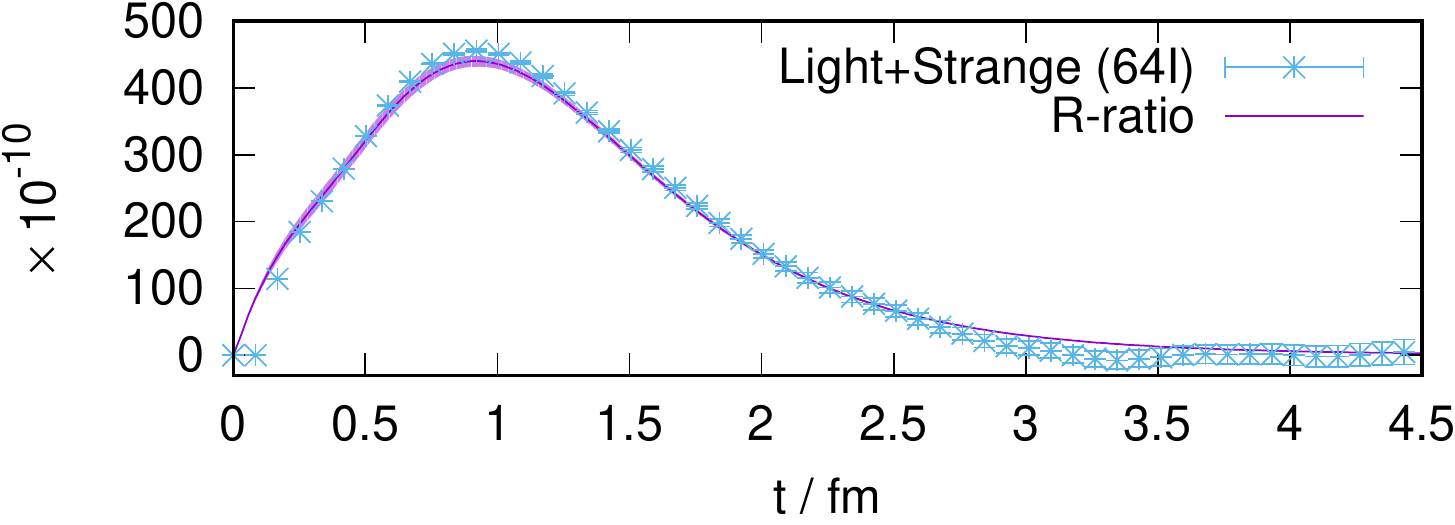}
      \caption{Comparison of $w_tC(t)$ obtained using R-ratio data
        \cite{Jegerlehner2017} and lattice data on our 64I ensemble.}
  \label{fig:compc0}
\end{figure}
In Fig.~\ref{fig:compc0} we compare a lattice and R-ratio evaluation
of $w_tC(t)$ and note that the R-ratio data is most precise at very short
and long distances, while the lattice data is most precise at
intermediate distances.  We are therefore led to also investigate a 
position-space ``window method'' \cite{Bernecker:2011gh,Lehner:2017kuc} 
and write
\begin{align}
  a_\mu = a_\mu^{\rm SD} + a_\mu^{\rm W} + a_\mu^{\rm LD} 
\end{align}
with $a_\mu^{\rm SD} = \sum_t C(t) w_t [1 - \Theta(t,t_0,\Delta)]$,
$a_\mu^{\rm W} = \sum_t C(t) w_t [ \Theta(t,t_0,\Delta) - \Theta(t,t_1,\Delta) ]$, and
$a_\mu^{\rm LD}= \sum_t C(t) w_t \Theta(t,t_1,\Delta)$, 
where each contribution is accessible from both lattice and R-ratio
data.  We define $\Theta(t,t',\Delta) = \left[1 + \tanh\left[ (t-t')
    / \Delta \right]\right]/2$ which we find to be helpful to control
the effect of discretization errors by the smearing parameter
$\Delta$.  We then take $a_\mu^{\rm SD}$ and $a_\mu^{\rm LD}$ from the
R-ratio data and $a_\mu^{\rm W}$ from the lattice.  In this work we
use $\Delta=0.15$ fm, which we find to provide a sufficiently sharp
transition without increasing discretization errors noticeably.  This
method takes the most precise
regions of both datasets and therefore may be a promising alternative
to the proposal of Ref.~\cite{Charles:2017snx}.

\section{Analysis and results}
In Tab.~\ref{tab:res} we show our results for the individual as well
as summed contributions to $a_\mu$ for the window method as well as a
pure lattice determination.  We quote statistical uncertainties for
the lattice data (S) and the R-ratio data (RST) separately.
\begin{table*}[tb]
  \centering
  \begin{tabular}{L{2.3cm}rL{6.6cm}rL{6.7cm}}\hline\hline
    $a_\mu^{\rm~ud,~conn,~isospin}$   &    $202.9$&$ \hspace{-2\tabcolsep}(1.4)_{\rm S}(0.2)_{\rm C}(0.1)_{\rm V}(0.2)_{\rm A}(0.2)_{\rm Z}$ &   $649.7$&$ \hspace{-2\tabcolsep}(14.2)_{\rm S}(2.8)_{\rm C}(3.7)_{\rm V}(1.5)_{\rm A}(0.4)_{\rm Z}(0.1)_{\rm E48}(0.1)_{\rm E64}$  \\
$a_\mu^{\rm~s,~conn,~isospin}$   &    $27.0$&$ \hspace{-2\tabcolsep}(0.2)_{\rm S}(0.0)_{\rm C}(0.1)_{\rm A}(0.0)_{\rm Z}$ &   $53.2$&$ \hspace{-2\tabcolsep}(0.4)_{\rm S}(0.0)_{\rm C}(0.3)_{\rm A}(0.0)_{\rm Z}$  \\
$a_\mu^{\rm~c,~conn,~isospin}$   &    $3.0$&$ \hspace{-2\tabcolsep}(0.0)_{\rm S}(0.1)_{\rm C}(0.0)_{\rm Z}(0.0)_{\rm M}$ &   $14.3$&$ \hspace{-2\tabcolsep}(0.0)_{\rm S}(0.7)_{\rm C}(0.1)_{\rm Z}(0.0)_{\rm M}$  \\
$a_\mu^{\rm~uds,~disc,~isospin}$   &    $-1.0$&$ \hspace{-2\tabcolsep}(0.1)_{\rm S}(0.0)_{\rm C}(0.0)_{\rm V}(0.0)_{\rm A}(0.0)_{\rm Z}$ &   $-11.2$&$ \hspace{-2\tabcolsep}(3.3)_{\rm S}(0.4)_{\rm V}(2.3)_{\rm L}$  \\
$a_\mu^{\rm~QED,~conn}$   &    $0.2$&$ \hspace{-2\tabcolsep}(0.2)_{\rm S}(0.0)_{\rm C}(0.0)_{\rm V}(0.0)_{\rm A}(0.0)_{\rm Z}(0.0)_{\rm E}$ &   $5.9$&$ \hspace{-2\tabcolsep}(5.7)_{\rm S}(0.3)_{\rm C}(1.2)_{\rm V}(0.0)_{\rm A}(0.0)_{\rm Z}(1.1)_{\rm E}$  \\
$a_\mu^{\rm~QED,~disc}$   &    $-0.2$&$ \hspace{-2\tabcolsep}(0.1)_{\rm S}(0.0)_{\rm C}(0.0)_{\rm V}(0.0)_{\rm A}(0.0)_{\rm Z}(0.0)_{\rm E}$ &   $-6.9$&$ \hspace{-2\tabcolsep}(2.1)_{\rm S}(0.4)_{\rm C}(1.4)_{\rm V}(0.0)_{\rm A}(0.0)_{\rm Z}(1.3)_{\rm E}$  \\
$a_\mu^{\rm~SIB}$   &    $0.1$&$ \hspace{-2\tabcolsep}(0.2)_{\rm S}(0.0)_{\rm C}(0.2)_{\rm V}(0.0)_{\rm A}(0.0)_{\rm Z}(0.0)_{\rm E48}$ &   $10.6$&$ \hspace{-2\tabcolsep}(4.3)_{\rm S}(0.6)_{\rm C}(6.6)_{\rm V}(0.1)_{\rm A}(0.0)_{\rm Z}(1.3)_{\rm E48}$  \\
\hline$a_\mu^{\rm~udsc,~isospin}$   &    $231.9$&$ \hspace{-2\tabcolsep}(1.4)_{\rm S}(0.2)_{\rm C}(0.1)_{\rm V}(0.3)_{\rm A}(0.2)_{\rm Z}(0.0)_{\rm M}$ &   $705.9$&$ \hspace{-2\tabcolsep}(14.6)_{\rm S}(2.9)_{\rm C}(3.7)_{\rm V}(1.8)_{\rm A}(0.4)_{\rm Z}(2.3)_{\rm L}(0.1)_{\rm E48}$ \\
 & && &$ \hspace{-2\tabcolsep}(0.1)_{\rm E64}(0.0)_{\rm M}$  \\
$a_\mu^{\rm~QED,~SIB}$   &    $0.1$&$ \hspace{-2\tabcolsep}(0.3)_{\rm S}(0.0)_{\rm C}(0.2)_{\rm V}(0.0)_{\rm A}(0.0)_{\rm Z}(0.0)_{\rm E}(0.0)_{\rm E48}$ &   $9.5$&$ \hspace{-2\tabcolsep}(7.4)_{\rm S}(0.7)_{\rm C}(6.9)_{\rm V}(0.1)_{\rm A}(0.0)_{\rm Z}(1.7)_{\rm E}(1.3)_{\rm E48}$  \\
$a_\mu^{\rm~R-ratio}$   &    $460.4$&$ \hspace{-2\tabcolsep}(0.7)_{\rm RST}(2.1)_{\rm RSY}$ && \\
\hline$a_\mu$   &    $692.5$&$ \hspace{-2\tabcolsep}(1.4)_{\rm S}(0.2)_{\rm C}(0.2)_{\rm V}(0.3)_{\rm A}(0.2)_{\rm Z}(0.0)_{\rm E}(0.0)_{\rm E48}$&   $715.4$&$ \hspace{-2\tabcolsep}(16.3)_{\rm S}(3.0)_{\rm C}(7.8)_{\rm V}(1.9)_{\rm A}(0.4)_{\rm Z}(1.7)_{\rm E}(2.3)_{\rm L}$ \\
 &  &$ \hspace{-2\tabcolsep}(0.0)_{\rm b}(0.1)_{\rm c}(0.0)_{\rm \SQ}(0.0)_{\rm \EQ}(0.0)_{\rm M}(0.7)_{\rm RST}(2.1)_{\rm RSY}$ & &$ \hspace{-2\tabcolsep}(1.5)_{\rm E48}(0.1)_{\rm E64}(0.3)_{\rm b}(0.2)_{\rm c}(1.1)_{\rm \SQ}(0.3)_{\rm \EQ}(0.0)_{\rm M}$  \\

    \hline\hline
  \end{tabular}
  \caption{Individual and summed contributions to $a_\mu$ multiplied by $10^{10}$.  The left column
    lists results for the window method with $t_0=0.4$ fm and $t_1=1$ fm.  The right column shows results
    for the pure first-principles lattice calculation.  The
    respective uncertainties are defined in the main text.}
  \label{tab:res}
\end{table*}
For the quark-connected up, down, and strange contributions, the
computation is performed on two ensembles with inverse lattice spacing
$a^{-1}=1.730(4)$ GeV (48I) as well as $a^{-1}=2.359(7)$ GeV (64I) and
a continuum limit is taken.  The discretization error (C) is estimated
by taking the maximum of the squared measured ${\cal O}(a^2)$
correction as well as a simple $(a\Lambda)^4$ estimate, where we take
$\Lambda=400$ MeV.  We find the results on the 48I ensemble to differ
only a few per-cent from the continuum limit.  This holds for the full
lattice contribution as well as the window contributions considered in
this work.  For the quark-connected charm contribution additional
ensembles described in Ref.~\cite{Boyle:2017jwu} are used and the
maximum of the above and a $(am_c)^4$ estimate is taken as
discretization error.  The remaining contributions are small and only
computed on the 48I ensemble for which we take $(a\Lambda)^2$ as
estimate of discretization errors.

For the up and down quark-connected and disconnected contributions, we
correct finite-volume effects to leading order in finite-volume
position-space chiral perturbation theory \cite{Aubin:2015rzx}.  Note
that in our previous publication of the quark-disconnected contribution
\cite{Blum:2015you}, we added this finite-volume correction as an
uncertainty but did not shift the central value.  We take the largest
ratio of $p^6$ to $p^4$ corrections of Tab.~1 of
Ref.~\cite{Bijnens:2017esv} as systematic error estimate of neglected
finite-volume errors (V).  For the SIB correction
we also include the sizeable difference of the corresponding finite and
infinite-volume chiral perturbation theory calculation as
finite-volume uncertainty.
For the QED correction, we repeat the computation using an
infinite-volume photon (QED$_\infty$ \cite{Lehner:2015bga}) and
include the difference to the QED$_{\rm L}$ result as a finite-volume error.
Further details of the QED$_\infty$ procedure are provided
as supplementary material.

We furthermore propagate uncertainties of the lattice spacing (A) and
the renormalization factors $Z_V$ (Z).  For the quark-disconnected
contribution we adopt the additional long-distance error discussed in
Ref.~\cite{Blum:2015you} (L) and for the charm contribution we
propagate uncertainties from the global fit procedure \cite{Boyle:2017jwu} (M).  Systematic
errors of the R-ratio computation are taken from
Ref.~\cite{Jegerlehner2017} and quoted as (RSY).  The neglected bottom
quark (b) and charm sea quark (c) contributions as well as
effects of neglected QED ($\overline{\rm Q}$) and
  SIB ($\overline{\rm S}$) diagrams are estimated as
described in the previous section.

For the QED and SIB corrections, we assume dominance
of the low-lying $\pi\pi$ and $\pi\gamma$ states and
fit $C^{(1)}_{\rm QED}(t)$ as well as $C^{(1)}_{\Delta m_f}(t)$ to
$(c_1 + c_0 t) e^{-E t}$, where we vary $c_0$ and $c_1$ for fixed energy $E$.  The resulting
p-values are larger than $0.2$ for all cases and we use
this functional form to compute the respective
contribution to $a_\mu$.  For the QED correction, we vary the energy
$E$ between the lowest $\pi\pi$ and $\pi\gamma$ energies and quote the
difference as additional uncertainty (E).  For the SIB correction, we
take $E$ to be the $\pi\pi$ ground-state energy.

For the light quark contribution of our pure lattice result we use a
bounding method \cite{LehnerTalkLGT16} similar to
Ref.~\cite{Borsanyi:2016lpl} and find that upper and lower bounds meet
within errors at $t=3.0$ fm.  We vary the ground-state energy that
enters this method \cite{Borsanyi:2017zdw} between the free-field and
interacting value \cite{Luscher:1986pf}.  For the 48I ensemble we find
$E_0^{\rm free} = 527.3$ MeV, $E_0 = 517.4$ MeV $+~{\cal O}(1/L^6)$
and for the 64I ensemble we have $E_0^{\rm free} = 536.1$ MeV, $E_0 =
525.1$ MeV $+~{\cal O}(1/L^6)$.  We quote the respective uncertainties
as (E48) and (E64).  The variation of $\pi\pi$ ground-state energy on
the 48I ensemble also enters the SIB correction as described above.

\begin{figure}[tb]
\begin{center}
\includegraphics[scale=0.55]{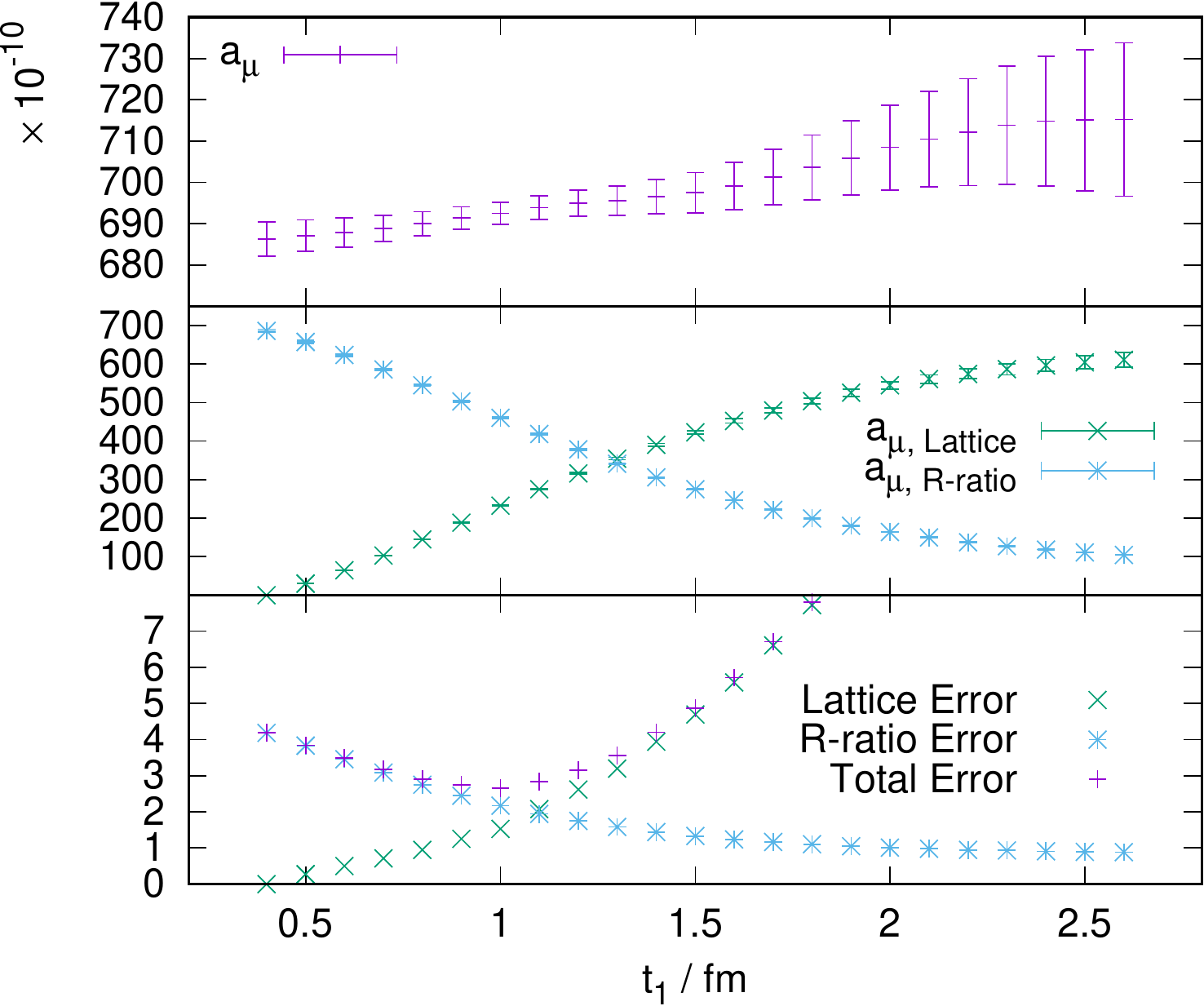}
\end{center} 
\caption{We show results for the window method with $t_0=0.4$ fm as a
  function of $t_1$.  The top panel shows the combined $a_\mu$, the
  middle panel shows the partial contributions of the lattice and
  R-ratio data, and the bottom shows the respective uncertainties.}
\label{fig:combi}
\end{figure}

\begin{figure}[tb]
\begin{center}
\includegraphics[scale=0.55]{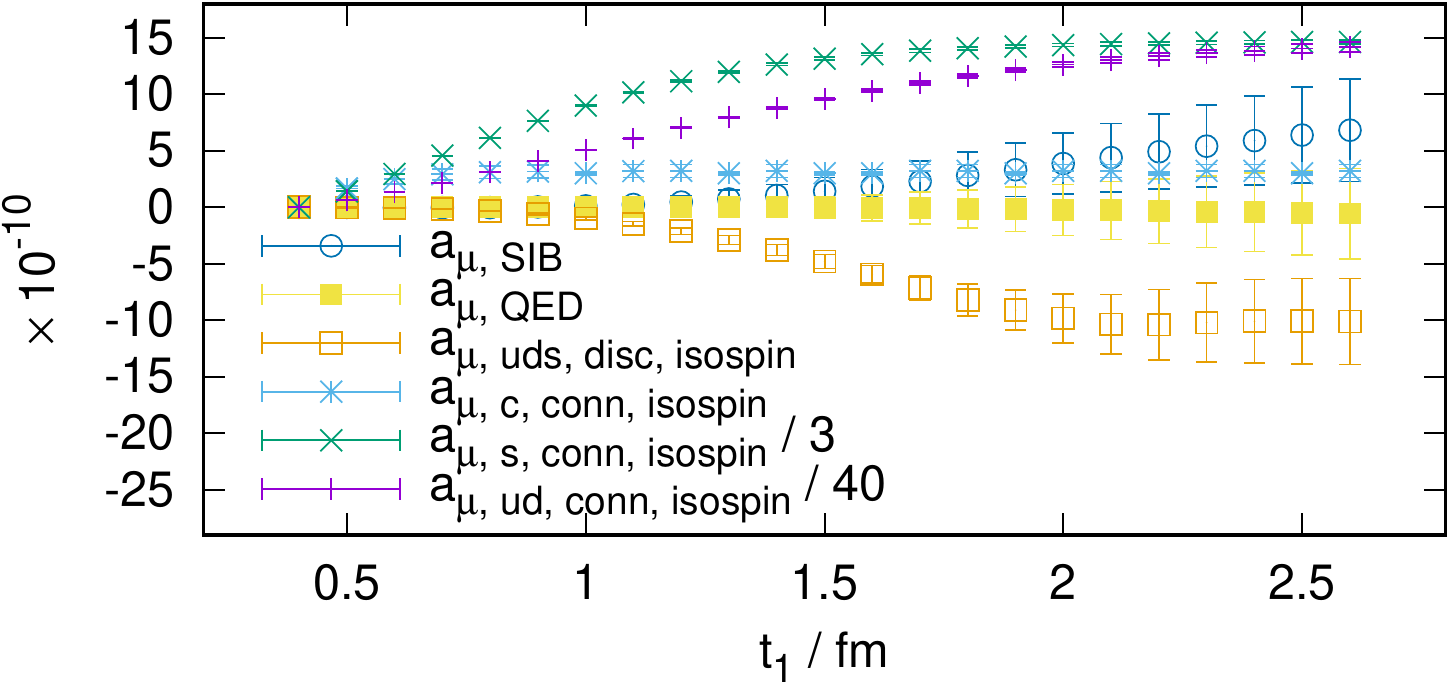}
\end{center} 
\caption{The individual lattice components of the window method with
  $t_0=0.4$ fm as function of $t_1$.}
\label{fig:contri}
\end{figure}

\begin{figure}[tb]
\begin{center}
\includegraphics[scale=0.55]{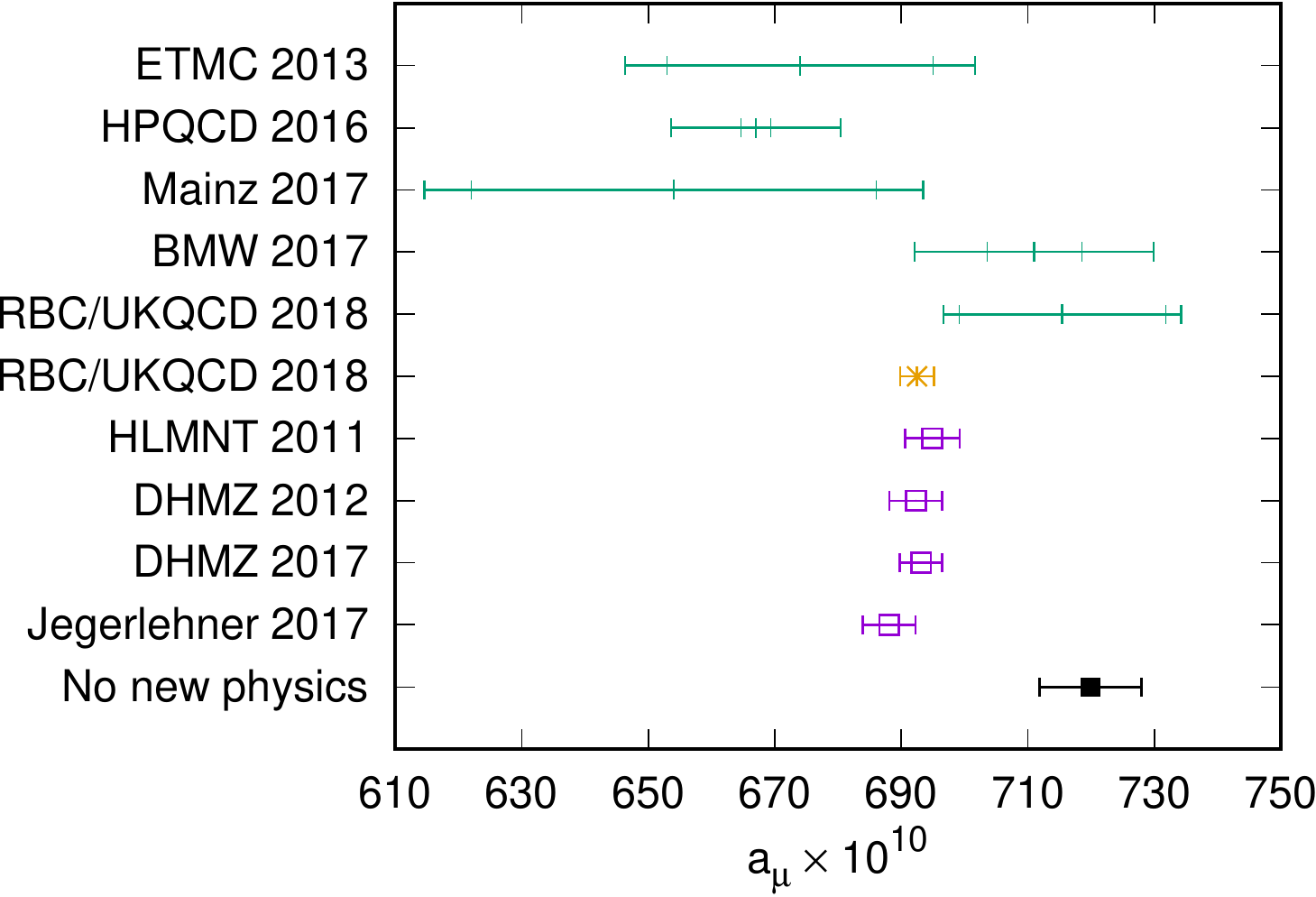}
\end{center} 
\caption{Our results (RBC/UKQCD 2018) compared to previously published
  results.  The green data-points are pure lattice computations, the
  orange data-point is our combined window analysis, and the purple
  data-points are pure R-ratio results.  The references are ETMC 2013
  \cite{Burger:2013jya}, HPQCD 2016 \cite{Chakraborty:2016mwy}, Mainz
  2017 \cite{DellaMorte:2017dyu}, BMW 2017 \cite{Borsanyi:2017zdw},
  HLMNT 2011 \cite{Hagiwara:2011af}, DHMZ 2012 \cite{Davier:2010nc},
  DHMZ 2017 \cite{Davier:2017zfy}, Jegerlehner 2017
  \cite{Jegerlehner:2017lbd}, and No new physics \cite{PDG2017}.  The
  innermost error-bar corresponds to the statistical uncertainty.}

\label{fig:overview}
\end{figure}

Figure~\ref{fig:combi} shows our results for the window method with
$t_0=0.4$ fm.  While the partial lattice and R-ratio contributions
change by several $100 \times 10^{-10}$, the sum changes only at the
level of quoted uncertainties.  This provides a non-trivial
consistency check between the lattice and the R-ratio data for length
scales between $0.4$ fm and $2.6$ fm. We expand on this check in the
supplementary material.  The uncertainty of the current analysis is
minimal for $t_1=1$ fm, which we take as our main result for the
window method.  For $t_0=t_1$ we reproduce the value of
Ref.~\cite{Jegerlehner2017}.  In Fig.~\ref{fig:contri}, we show the
$t_1$-dependence of individual lattice contributions and compare our
results with previously published results in Fig.~\ref{fig:overview}.
Our combined lattice and R-ratio result is more precise than
  the R-ratio computation by itself and reduces the tension to the
  other R-ratio results.  Results for different window parameters
$t_0$ and $t_1$ and a comparison of individual components with
previously published results are provided as supplementary material.

\section{Conclusion}
We have presented both a complete first-principles calculation of the
leading-order hadronic vacuum polarization contribution to the muon
anomalous magnetic moment from lattice QCD+QED at physical pion mass
as well as a combination with R-ratio data.  For the former we find
$a_\mu^{\rm HVP~LO}=715.4(16.3)(9.2) \times 10^{-10}$, where the first
error is statistical and the second is systematic.
For the latter we find $a_\mu^{\rm HVP~LO} = 692.5(1.4)(0.5)(0.7)(2.1)
\times 10^{-10}$ with lattice statistical, lattice systematic, R-ratio
statistical, and R-ratio systematic errors given separately.  This is
the currently most precise determination of $a_\mu^{\rm HVP~LO}$
corresponding to a $3.7\sigma$ tension
\begin{align}
  a_\mu^{\rm EXP} - a_\mu^{\rm SM} = \,\,&27.4(2.7)(2.6)(6.3)  \times 10 ^ {-10} \,.
\end{align}
The presented combination of lattice and R-ratio data also serves to
provide additional non-trivial cross-checks between lattice and
R-ratio data.  The precision of
this computation will be improved in future work including
simulations at smaller lattice spacings and at larger volumes.

\section{Acknowledgments}
We would like to thank our RBC and UKQCD collaborators for helpful
discussions and support.  We would also like to thank Kim Maltman and
Masashi Hayakawa for valuable discussions.  We are indebted to Fred
Jegerlehner for helpful exchanges on the R-ratio compilation of
Ref.~\cite{Jegerlehner2017}.  P.A.B., A.P., and J.T.T. are
  supported in part by UK STFC grants ST/L000458/1 and
  ST/P000711/1. A.P. also received funding from the European Research
  Council (ERC) under the European Union's Horizon 2020 research and
  innovation programme under grant agreement No 757646.  T.B. is
supported by U.S. Department of Energy Grant No. DE-FG02-92ER40716.
V.G. and A.J. acknowledge support from STFC consolidated grant
ST/P000711/1, A.J. has also received funding from the European
Research Council under the EU FP7 Programme (FP7/2007-2013) / ERC
Grant agreement 279757.  T.I., C.J., and C.L. are supported in part by
US DOE Contract DESC0012704(BNL).  T.I. is also supported by JSPS
KAKENHI grant numbers JP26400261, JP17H02906 and by MEXT as ``Priority
Issue on Post-K computer'' (Elucidation of the Fundamental Laws and
Evolution of the Universe) and JICFuS.  C.L. is also supported by a
DOE Office of Science Early Career Award.  L.J. is supported by the
Department of Energy, Laboratory Directed Research and Development
(LDRD) funding of BNL, under contract DE-EC0012704.  This work was
supported by resources provided by the Scientific Data and Computing
Center (SDCC) at Brookhaven National Laboratory (BNL), a DOE Office of
Science User Facility supported by the Office of Science of the US
Department of Energy.  The SDCC is a major component of the
Computational Science Initiative at BNL.  We gratefully acknowledge
computing resources provided through USQCD clusters at Fermilab and
Jefferson Lab as well as the IBM Blue Gene/Q (BG/Q) Mira machine at
the Argonne Leadership Class Facility, a DOE Office of Science
Facility supported under Contract De-AC02-06CH11357.  This work was
also supported by the DiRAC Blue Gene Q Shared Petaflop system at the
University of Edinburgh, operated by the Edinburgh Parallel Computing
Centre on behalf of the STFC DiRAC HPC Facility (www.dirac.ac.uk).
This equipment was funded by BIS National E-infrastructure capital
grant ST/K000411/1, STFC capital grant ST/H008845/1, and STFC DiRAC
Operations grants ST/K005804/1 and ST/K005790/1.  DiRAC is part of the
National E-Infrastructure.  The software used includes
\href{http://www2.ph.ed.ac.uk/~paboyle/bagel/Bagel.html}{BAGEL} (GNU
GPLv2 license), \href{http://qcdoc.phys.columbia.edu/cps.html}{CPS},
\href{http://cuths01.phys.columbia.edu:8080/christoph/gpt}{GPT},
\href{https://github.com/paboyle/Grid/}{Grid} (GNU GPLv2 license), and
\href{https://github.com/aportelli/LatAnalyze}{LatAnalyze} (GNU GPLv3
license).

\bibliography{refs}

\clearpage

\setcounter{page}{1}
\renewcommand{\thepage}{Supplementary Information -- S\arabic{page}}
\setcounter{table}{0}
\renewcommand{\thetable}{S\,\Roman{table}}
\setcounter{equation}{0}
\renewcommand{\theequation}{S\,\arabic{equation}}

\section{Supplementary Material}
In this section we expand on a selection of technical details and add
results to facilitate cross-checks of different calculations of
$a_\mu^{\rm HVP~LO}$.
\begin{table*}[tbp]
  \centering
  \begin{tabular}{L{3.3cm}rL{12.9cm}}\hline\hline
    $a_\mu^{\rm~ud,~conn,~isospin}$  &   $317.8($& \hspace{-2\tabcolsep}$2.9)_{\rm S}(0.2)_{\rm C}(0.4)_{\rm V}(0.7)_{\rm A}(0.3)_{\rm Z}$ \\
$a_\mu^{\rm~s,~conn,~isospin}$  &   $35.9($& \hspace{-2\tabcolsep}$0.2)_{\rm S}(0.0)_{\rm C}(0.2)_{\rm A}(0.0)_{\rm Z}$ \\
$a_\mu^{\rm~c,~conn,~isospin}$  &   $3.0($& \hspace{-2\tabcolsep}$0.0)_{\rm S}(0.1)_{\rm C}(0.0)_{\rm Z}(0.0)_{\rm M}$ \\
$a_\mu^{\rm~uds,~disc,~isospin}$  &   $-2.9($& \hspace{-2\tabcolsep}$0.4)_{\rm S}(0.1)_{\rm C}(0.0)_{\rm V}(0.0)_{\rm A}(0.0)_{\rm Z}$ \\
$a_\mu^{\rm~QED,~conn}$  &   $0.6($& \hspace{-2\tabcolsep}$0.5)_{\rm S}(0.0)_{\rm C}(0.1)_{\rm V}(0.0)_{\rm A}(0.0)_{\rm Z}(0.0)_{\rm E}$ \\
$a_\mu^{\rm~QED,~disc}$  &   $-0.6($& \hspace{-2\tabcolsep}$0.2)_{\rm S}(0.0)_{\rm C}(0.1)_{\rm V}(0.0)_{\rm A}(0.0)_{\rm Z}(0.0)_{\rm E}$ \\
$a_\mu^{\rm~SIB}$  &   $0.7($& \hspace{-2\tabcolsep}$0.6)_{\rm S}(0.0)_{\rm C}(0.5)_{\rm V}(0.0)_{\rm A}(0.0)_{\rm Z}(0.1)_{\rm E48}$ \\
\hline
$a_\mu^{\rm~udsc,~isospin}$  &   $353.9($& \hspace{-2\tabcolsep}$2.9)_{\rm S}(0.3)_{\rm C}(0.4)_{\rm V}(0.8)_{\rm A}(0.3)_{\rm Z}(0.0)_{\rm M}$ \\
$a_\mu^{\rm~QED,~SIB}$  &   $0.7($& \hspace{-2\tabcolsep}$0.8)_{\rm S}(0.1)_{\rm C}(0.5)_{\rm V}(0.0)_{\rm A}(0.0)_{\rm Z}(0.0)_{\rm E}(0.1)_{\rm E48}$ \\
$a_\mu^{\rm~R-ratio}$  &   $340.9($& \hspace{-2\tabcolsep}$0.5)_{\rm RST}(1.5)_{\rm RSY}$ \\
\hline
$a_\mu$  &   $695.5($& \hspace{-2\tabcolsep}$3.0)_{\rm S}(0.3)_{\rm C}(0.6)_{\rm V}(0.8)_{\rm A}(0.3)_{\rm Z}(0.0)_{\rm E}(0.0)_{\rm M}(0.1)_{\rm E48}(0.0)_{\rm b}(0.1)_{\rm c}(0.1)_{\rm \SQ}(0.0)_{\rm \EQ}(0.5)_{\rm RST}(1.5)_{\rm RSY}$ \\

    \hline\hline
  \end{tabular}
  \caption{Individual and summed contributions to $a_\mu$ multiplied by $10^{10}$ for the window method with $t_0=0.4$ fm and $t_1=1.3$ fm.  The
    respective uncertainties are defined in the main text.}
  \label{tab:addtab0}
\end{table*}

\begin{table*}[tbp]
  \centering
  \begin{tabular}{L{3.3cm}rL{12.9cm}}\hline\hline
    $a_\mu^{\rm~ud,~conn,~isospin}$  &   $413.1($& \hspace{-2\tabcolsep}$4.9)_{\rm S}(0.3)_{\rm C}(0.7)_{\rm V}(1.3)_{\rm A}(0.3)_{\rm Z}$ \\
$a_\mu^{\rm~s,~conn,~isospin}$  &   $40.6($& \hspace{-2\tabcolsep}$0.3)_{\rm S}(0.0)_{\rm C}(0.2)_{\rm A}(0.0)_{\rm Z}$ \\
$a_\mu^{\rm~c,~conn,~isospin}$  &   $3.0($& \hspace{-2\tabcolsep}$0.0)_{\rm S}(0.1)_{\rm C}(0.0)_{\rm Z}(0.0)_{\rm M}$ \\
$a_\mu^{\rm~uds,~disc,~isospin}$  &   $-5.9($& \hspace{-2\tabcolsep}$0.7)_{\rm S}(0.3)_{\rm C}(0.1)_{\rm V}(0.0)_{\rm A}(0.0)_{\rm Z}$ \\
$a_\mu^{\rm~QED,~conn}$  &   $1.3($& \hspace{-2\tabcolsep}$1.0)_{\rm S}(0.1)_{\rm C}(0.3)_{\rm V}(0.0)_{\rm A}(0.0)_{\rm Z}(0.1)_{\rm E}$ \\
$a_\mu^{\rm~QED,~disc}$  &   $-1.3($& \hspace{-2\tabcolsep}$0.4)_{\rm S}(0.1)_{\rm C}(0.3)_{\rm V}(0.0)_{\rm A}(0.0)_{\rm Z}(0.1)_{\rm E}$ \\
$a_\mu^{\rm~SIB}$  &   $1.9($& \hspace{-2\tabcolsep}$1.1)_{\rm S}(0.1)_{\rm C}(0.9)_{\rm V}(0.0)_{\rm A}(0.0)_{\rm Z}(0.2)_{\rm E48}$ \\
\hline
$a_\mu^{\rm~udsc,~isospin}$  &   $450.8($& \hspace{-2\tabcolsep}$5.0)_{\rm S}(0.5)_{\rm C}(0.7)_{\rm V}(1.5)_{\rm A}(0.4)_{\rm Z}(0.0)_{\rm M}$ \\
$a_\mu^{\rm~QED,~SIB}$  &   $1.8($& \hspace{-2\tabcolsep}$1.6)_{\rm S}(0.2)_{\rm C}(1.0)_{\rm V}(0.0)_{\rm A}(0.0)_{\rm Z}(0.1)_{\rm E}(0.2)_{\rm E48}$ \\
$a_\mu^{\rm~R-ratio}$  &   $246.5($& \hspace{-2\tabcolsep}$0.5)_{\rm RST}(1.2)_{\rm RSY}$ \\
\hline
$a_\mu$  &   $699.1($& \hspace{-2\tabcolsep}$5.2)_{\rm S}(0.5)_{\rm C}(1.2)_{\rm V}(1.5)_{\rm A}(0.4)_{\rm Z}(0.1)_{\rm E}(0.0)_{\rm M}(0.2)_{\rm E48}(0.0)_{\rm b}(0.1)_{\rm c}(0.2)_{\rm \SQ}(0.0)_{\rm \EQ}(0.5)_{\rm RST}(1.2)_{\rm RSY}$ \\

    \hline\hline
  \end{tabular}
  \caption{Individual and summed contributions to $a_\mu$ multiplied by $10^{10}$ for the window method with $t_0=0.4$ fm and $t_1=1.6$ fm.  The
    respective uncertainties are defined in the main text.}
  \label{tab:addtab1}
\end{table*}

\begin{table*}[tbp]
  \centering
  \begin{tabular}{L{3.3cm}rL{12.9cm}}\hline\hline
    $a_\mu^{\rm~ud,~conn,~isospin}$  &   $485.7($& \hspace{-2\tabcolsep}$7.8)_{\rm S}(0.4)_{\rm C}(1.1)_{\rm V}(1.9)_{\rm A}(0.4)_{\rm Z}$ \\
$a_\mu^{\rm~s,~conn,~isospin}$  &   $42.7($& \hspace{-2\tabcolsep}$0.4)_{\rm S}(0.0)_{\rm C}(0.2)_{\rm A}(0.0)_{\rm Z}$ \\
$a_\mu^{\rm~c,~conn,~isospin}$  &   $3.0($& \hspace{-2\tabcolsep}$0.0)_{\rm S}(0.1)_{\rm C}(0.0)_{\rm Z}(0.0)_{\rm M}$ \\
$a_\mu^{\rm~uds,~disc,~isospin}$  &   $-9.1($& \hspace{-2\tabcolsep}$1.7)_{\rm S}(0.4)_{\rm C}(0.1)_{\rm V}(0.0)_{\rm A}(0.0)_{\rm Z}$ \\
$a_\mu^{\rm~QED,~conn}$  &   $2.0($& \hspace{-2\tabcolsep}$1.7)_{\rm S}(0.1)_{\rm C}(0.4)_{\rm V}(0.0)_{\rm A}(0.0)_{\rm Z}(0.2)_{\rm E}$ \\
$a_\mu^{\rm~QED,~disc}$  &   $-2.2($& \hspace{-2\tabcolsep}$0.6)_{\rm S}(0.2)_{\rm C}(0.4)_{\rm V}(0.0)_{\rm A}(0.0)_{\rm Z}(0.2)_{\rm E}$ \\
$a_\mu^{\rm~SIB}$  &   $3.3($& \hspace{-2\tabcolsep}$1.7)_{\rm S}(0.2)_{\rm C}(1.6)_{\rm V}(0.0)_{\rm A}(0.0)_{\rm Z}(0.3)_{\rm E48}$ \\
\hline
$a_\mu^{\rm~udsc,~isospin}$  &   $522.3($& \hspace{-2\tabcolsep}$8.0)_{\rm S}(0.6)_{\rm C}(1.1)_{\rm V}(2.1)_{\rm A}(0.4)_{\rm Z}(0.0)_{\rm M}$ \\
$a_\mu^{\rm~QED,~SIB}$  &   $3.1($& \hspace{-2\tabcolsep}$2.5)_{\rm S}(0.3)_{\rm C}(1.7)_{\rm V}(0.0)_{\rm A}(0.0)_{\rm Z}(0.2)_{\rm E}(0.3)_{\rm E48}$ \\
$a_\mu^{\rm~R-ratio}$  &   $180.5($& \hspace{-2\tabcolsep}$0.4)_{\rm RST}(1.0)_{\rm RSY}$ \\
\hline
$a_\mu$  &   $705.9($& \hspace{-2\tabcolsep}$8.4)_{\rm S}(0.6)_{\rm C}(2.0)_{\rm V}(2.1)_{\rm A}(0.4)_{\rm Z}(0.2)_{\rm E}(0.0)_{\rm M}(0.3)_{\rm E48}(0.0)_{\rm b}(0.1)_{\rm c}(0.3)_{\rm \SQ}(0.1)_{\rm \EQ}(0.4)_{\rm RST}(1.0)_{\rm RSY}$ \\

    \hline\hline
  \end{tabular}
  \caption{Individual and summed contributions to $a_\mu$ multiplied by $10^{10}$ for the window method with $t_0=0.4$ fm and $t_1=1.9$ fm.  The
    respective uncertainties are defined in the main text.}
  \label{tab:addtab2}
\end{table*}

\begin{table*}[tbp]
  \centering
  \begin{tabular}{L{3.3cm}rL{12.9cm}}\hline\hline
    $a_\mu^{\rm~ud,~conn,~isospin}$  &   $533.8($& \hspace{-2\tabcolsep}$11.3)_{\rm S}(0.4)_{\rm C}(1.6)_{\rm V}(2.3)_{\rm A}(0.4)_{\rm Z}$ \\
$a_\mu^{\rm~s,~conn,~isospin}$  &   $43.5($& \hspace{-2\tabcolsep}$0.4)_{\rm S}(0.0)_{\rm C}(0.3)_{\rm A}(0.0)_{\rm Z}$ \\
$a_\mu^{\rm~c,~conn,~isospin}$  &   $3.0($& \hspace{-2\tabcolsep}$0.0)_{\rm S}(0.1)_{\rm C}(0.0)_{\rm Z}(0.0)_{\rm M}$ \\
$a_\mu^{\rm~uds,~disc,~isospin}$  &   $-10.4($& \hspace{-2\tabcolsep}$3.1)_{\rm S}(0.5)_{\rm C}(0.2)_{\rm V}(0.1)_{\rm A}(0.0)_{\rm Z}$ \\
$a_\mu^{\rm~QED,~conn}$  &   $2.8($& \hspace{-2\tabcolsep}$2.5)_{\rm S}(0.2)_{\rm C}(0.6)_{\rm V}(0.0)_{\rm A}(0.0)_{\rm Z}(0.3)_{\rm E}$ \\
$a_\mu^{\rm~QED,~disc}$  &   $-3.2($& \hspace{-2\tabcolsep}$0.9)_{\rm S}(0.2)_{\rm C}(0.6)_{\rm V}(0.0)_{\rm A}(0.0)_{\rm Z}(0.3)_{\rm E}$ \\
$a_\mu^{\rm~SIB}$  &   $4.9($& \hspace{-2\tabcolsep}$2.3)_{\rm S}(0.2)_{\rm C}(2.3)_{\rm V}(0.1)_{\rm A}(0.0)_{\rm Z}(0.5)_{\rm E48}$ \\
\hline
$a_\mu^{\rm~udsc,~isospin}$  &   $570.0($& \hspace{-2\tabcolsep}$11.7)_{\rm S}(0.6)_{\rm C}(1.6)_{\rm V}(2.6)_{\rm A}(0.5)_{\rm Z}(0.0)_{\rm M}$ \\
$a_\mu^{\rm~QED,~SIB}$  &   $4.6($& \hspace{-2\tabcolsep}$3.6)_{\rm S}(0.4)_{\rm C}(2.5)_{\rm V}(0.1)_{\rm A}(0.0)_{\rm Z}(0.4)_{\rm E}(0.5)_{\rm E48}$ \\
$a_\mu^{\rm~R-ratio}$  &   $137.6($& \hspace{-2\tabcolsep}$0.4)_{\rm RST}(0.9)_{\rm RSY}$ \\
\hline
$a_\mu$  &   $712.2($& \hspace{-2\tabcolsep}$12.2)_{\rm S}(0.8)_{\rm C}(2.9)_{\rm V}(2.6)_{\rm A}(0.4)_{\rm Z}(0.4)_{\rm E}(0.0)_{\rm M}(0.5)_{\rm E48}(0.0)_{\rm b}(0.1)_{\rm c}(0.5)_{\rm \SQ}(0.1)_{\rm \EQ}(0.4)_{\rm RST}(0.9)_{\rm RSY}$ \\

    \hline\hline
  \end{tabular}
  \caption{Individual and summed contributions to $a_\mu$ multiplied by $10^{10}$ for the window method with $t_0=0.4$ fm and $t_1=2.2$ fm.  The
    respective uncertainties are defined in the main text.}
  \label{tab:addtab3}
\end{table*}

\begin{table*}[tbp]
  \centering
  \begin{tabular}{L{3.3cm}rL{12.9cm}}\hline\hline
    $a_\mu^{\rm~ud,~conn,~isospin}$  &   $561.6($& \hspace{-2\tabcolsep}$15.2)_{\rm S}(0.4)_{\rm C}(2.0)_{\rm V}(2.5)_{\rm A}(0.4)_{\rm Z}$ \\
$a_\mu^{\rm~s,~conn,~isospin}$  &   $43.8($& \hspace{-2\tabcolsep}$0.4)_{\rm S}(0.0)_{\rm C}(0.3)_{\rm A}(0.0)_{\rm Z}$ \\
$a_\mu^{\rm~c,~conn,~isospin}$  &   $3.0($& \hspace{-2\tabcolsep}$0.0)_{\rm S}(0.1)_{\rm C}(0.0)_{\rm Z}(0.0)_{\rm M}$ \\
$a_\mu^{\rm~uds,~disc,~isospin}$  &   $-10.1($& \hspace{-2\tabcolsep}$3.7)_{\rm S}(0.5)_{\rm C}(0.2)_{\rm V}(0.0)_{\rm A}(0.0)_{\rm Z}$ \\
$a_\mu^{\rm~QED,~conn}$  &   $3.6($& \hspace{-2\tabcolsep}$3.3)_{\rm S}(0.3)_{\rm C}(0.7)_{\rm V}(0.0)_{\rm A}(0.0)_{\rm Z}(0.4)_{\rm E}$ \\
$a_\mu^{\rm~QED,~disc}$  &   $-4.1($& \hspace{-2\tabcolsep}$1.2)_{\rm S}(0.3)_{\rm C}(0.8)_{\rm V}(0.0)_{\rm A}(0.0)_{\rm Z}(0.5)_{\rm E}$ \\
$a_\mu^{\rm~SIB}$  &   $6.4($& \hspace{-2\tabcolsep}$2.9)_{\rm S}(0.3)_{\rm C}(3.1)_{\rm V}(0.1)_{\rm A}(0.0)_{\rm Z}(0.7)_{\rm E48}$ \\
\hline
$a_\mu^{\rm~udsc,~isospin}$  &   $598.4($& \hspace{-2\tabcolsep}$15.7)_{\rm S}(0.6)_{\rm C}(2.0)_{\rm V}(2.8)_{\rm A}(0.5)_{\rm Z}(0.0)_{\rm M}$ \\
$a_\mu^{\rm~QED,~SIB}$  &   $5.9($& \hspace{-2\tabcolsep}$4.5)_{\rm S}(0.5)_{\rm C}(3.2)_{\rm V}(0.1)_{\rm A}(0.0)_{\rm Z}(0.7)_{\rm E}(0.7)_{\rm E48}$ \\
$a_\mu^{\rm~R-ratio}$  &   $110.8($& \hspace{-2\tabcolsep}$0.3)_{\rm RST}(0.8)_{\rm RSY}$ \\
\hline
$a_\mu$  &   $715.1($& \hspace{-2\tabcolsep}$16.3)_{\rm S}(0.8)_{\rm C}(3.8)_{\rm V}(2.9)_{\rm A}(0.5)_{\rm Z}(0.7)_{\rm E}(0.0)_{\rm M}(0.7)_{\rm E48}(0.0)_{\rm b}(0.1)_{\rm c}(0.6)_{\rm \SQ}(0.1)_{\rm \EQ}(0.3)_{\rm RST}(0.8)_{\rm RSY}$ \\

    \hline\hline
  \end{tabular}
  \caption{Individual and summed contributions to $a_\mu$ multiplied by $10^{10}$ for the window method with $t_0=0.4$ fm and $t_1=2.5$ fm.  The
    respective uncertainties are defined in the main text.}
  \label{tab:addtab4}
\end{table*}

\begin{table*}[tbp]
  \centering
  \begin{tabular}{L{3.3cm}rL{12.9cm}}\hline\hline
    $a_\mu^{\rm~ud,~conn,~isospin}$  &   $222.5($& \hspace{-2\tabcolsep}$1.5)_{\rm S}(0.2)_{\rm C}(0.1)_{\rm V}(0.2)_{\rm A}(0.2)_{\rm Z}$ \\
$a_\mu^{\rm~s,~conn,~isospin}$  &   $30.5($& \hspace{-2\tabcolsep}$0.2)_{\rm S}(0.0)_{\rm C}(0.1)_{\rm A}(0.0)_{\rm Z}$ \\
$a_\mu^{\rm~c,~conn,~isospin}$  &   $5.6($& \hspace{-2\tabcolsep}$0.0)_{\rm S}(0.3)_{\rm C}(0.0)_{\rm Z}(0.0)_{\rm M}$ \\
$a_\mu^{\rm~uds,~disc,~isospin}$  &   $-1.0($& \hspace{-2\tabcolsep}$0.1)_{\rm S}(0.0)_{\rm C}(0.0)_{\rm V}(0.0)_{\rm A}(0.0)_{\rm Z}$ \\
$a_\mu^{\rm~QED,~conn}$  &   $0.2($& \hspace{-2\tabcolsep}$0.2)_{\rm S}(0.0)_{\rm C}(0.0)_{\rm V}(0.0)_{\rm A}(0.0)_{\rm Z}(0.0)_{\rm E}$ \\
$a_\mu^{\rm~QED,~disc}$  &   $-0.2($& \hspace{-2\tabcolsep}$0.1)_{\rm S}(0.0)_{\rm C}(0.0)_{\rm V}(0.0)_{\rm A}(0.0)_{\rm Z}(0.0)_{\rm E}$ \\
$a_\mu^{\rm~SIB}$  &   $0.1($& \hspace{-2\tabcolsep}$0.2)_{\rm S}(0.0)_{\rm C}(0.2)_{\rm V}(0.0)_{\rm A}(0.0)_{\rm Z}(0.0)_{\rm E48}$ \\
\hline
$a_\mu^{\rm~udsc,~isospin}$  &   $257.6($& \hspace{-2\tabcolsep}$1.5)_{\rm S}(0.3)_{\rm C}(0.2)_{\rm V}(0.3)_{\rm A}(0.2)_{\rm Z}(0.0)_{\rm M}$ \\
$a_\mu^{\rm~QED,~SIB}$  &   $0.1($& \hspace{-2\tabcolsep}$0.3)_{\rm S}(0.0)_{\rm C}(0.2)_{\rm V}(0.0)_{\rm A}(0.0)_{\rm Z}(0.0)_{\rm E}(0.0)_{\rm E48}$ \\
$a_\mu^{\rm~R-ratio}$  &   $436.2($& \hspace{-2\tabcolsep}$0.6)_{\rm RST}(1.8)_{\rm RSY}$ \\
\hline
$a_\mu$  &   $694.0($& \hspace{-2\tabcolsep}$1.5)_{\rm S}(0.3)_{\rm C}(0.2)_{\rm V}(0.3)_{\rm A}(0.2)_{\rm Z}(0.0)_{\rm E}(0.0)_{\rm M}(0.0)_{\rm E48}(0.0)_{\rm b}(0.1)_{\rm c}(0.0)_{\rm \SQ}(0.0)_{\rm \EQ}(0.6)_{\rm RST}(1.8)_{\rm RSY}$ \\

    \hline\hline
  \end{tabular}
  \caption{Individual and summed contributions to $a_\mu$ multiplied by $10^{10}$ for the window method with $t_0=0.3$ fm and $t_1=1$ fm.  The
    respective uncertainties are defined in the main text.}
  \label{tab:addtab5}
\end{table*}

\begin{table*}[tbp]
  \centering
  \begin{tabular}{L{3.3cm}rL{12.9cm}}\hline\hline
    $a_\mu^{\rm~ud,~conn,~isospin}$  &   $178.8($& \hspace{-2\tabcolsep}$1.3)_{\rm S}(0.1)_{\rm C}(0.1)_{\rm V}(0.2)_{\rm A}(0.2)_{\rm Z}$ \\
$a_\mu^{\rm~s,~conn,~isospin}$  &   $22.8($& \hspace{-2\tabcolsep}$0.1)_{\rm S}(0.0)_{\rm C}(0.1)_{\rm A}(0.0)_{\rm Z}$ \\
$a_\mu^{\rm~c,~conn,~isospin}$  &   $1.4($& \hspace{-2\tabcolsep}$0.0)_{\rm S}(0.1)_{\rm C}(0.0)_{\rm Z}(0.0)_{\rm M}$ \\
$a_\mu^{\rm~uds,~disc,~isospin}$  &   $-1.0($& \hspace{-2\tabcolsep}$0.1)_{\rm S}(0.0)_{\rm C}(0.0)_{\rm V}(0.0)_{\rm A}(0.0)_{\rm Z}$ \\
$a_\mu^{\rm~QED,~conn}$  &   $0.2($& \hspace{-2\tabcolsep}$0.1)_{\rm S}(0.0)_{\rm C}(0.0)_{\rm V}(0.0)_{\rm A}(0.0)_{\rm Z}(0.0)_{\rm E}$ \\
$a_\mu^{\rm~QED,~disc}$  &   $-0.2($& \hspace{-2\tabcolsep}$0.1)_{\rm S}(0.0)_{\rm C}(0.0)_{\rm V}(0.0)_{\rm A}(0.0)_{\rm Z}(0.0)_{\rm E}$ \\
$a_\mu^{\rm~SIB}$  &   $0.1($& \hspace{-2\tabcolsep}$0.2)_{\rm S}(0.0)_{\rm C}(0.2)_{\rm V}(0.0)_{\rm A}(0.0)_{\rm Z}(0.0)_{\rm E48}$ \\
\hline
$a_\mu^{\rm~udsc,~isospin}$  &   $202.0($& \hspace{-2\tabcolsep}$1.3)_{\rm S}(0.2)_{\rm C}(0.1)_{\rm V}(0.3)_{\rm A}(0.2)_{\rm Z}(0.0)_{\rm M}$ \\
$a_\mu^{\rm~QED,~SIB}$  &   $0.2($& \hspace{-2\tabcolsep}$0.3)_{\rm S}(0.0)_{\rm C}(0.2)_{\rm V}(0.0)_{\rm A}(0.0)_{\rm Z}(0.0)_{\rm E}(0.0)_{\rm E48}$ \\
$a_\mu^{\rm~R-ratio}$  &   $489.5($& \hspace{-2\tabcolsep}$0.7)_{\rm RST}(2.4)_{\rm RSY}$ \\
\hline
$a_\mu$  &   $691.7($& \hspace{-2\tabcolsep}$1.4)_{\rm S}(0.2)_{\rm C}(0.2)_{\rm V}(0.3)_{\rm A}(0.2)_{\rm Z}(0.0)_{\rm E}(0.0)_{\rm M}(0.0)_{\rm E48}(0.0)_{\rm b}(0.1)_{\rm c}(0.0)_{\rm \SQ}(0.0)_{\rm \EQ}(0.7)_{\rm RST}(2.4)_{\rm RSY}$ \\

    \hline\hline
  \end{tabular}
  \caption{Individual and summed contributions to $a_\mu$ multiplied by $10^{10}$ for the window method with $t_0=0.5$ fm and $t_1=1$ fm.  The
    respective uncertainties are defined in the main text.}
  \label{tab:addtab6}
\end{table*}

\newcommand{\skp}{\vspace{0.4cm}}

\begin{figure}[b] 
  \begin{center}
    \includegraphics[width=7cm]{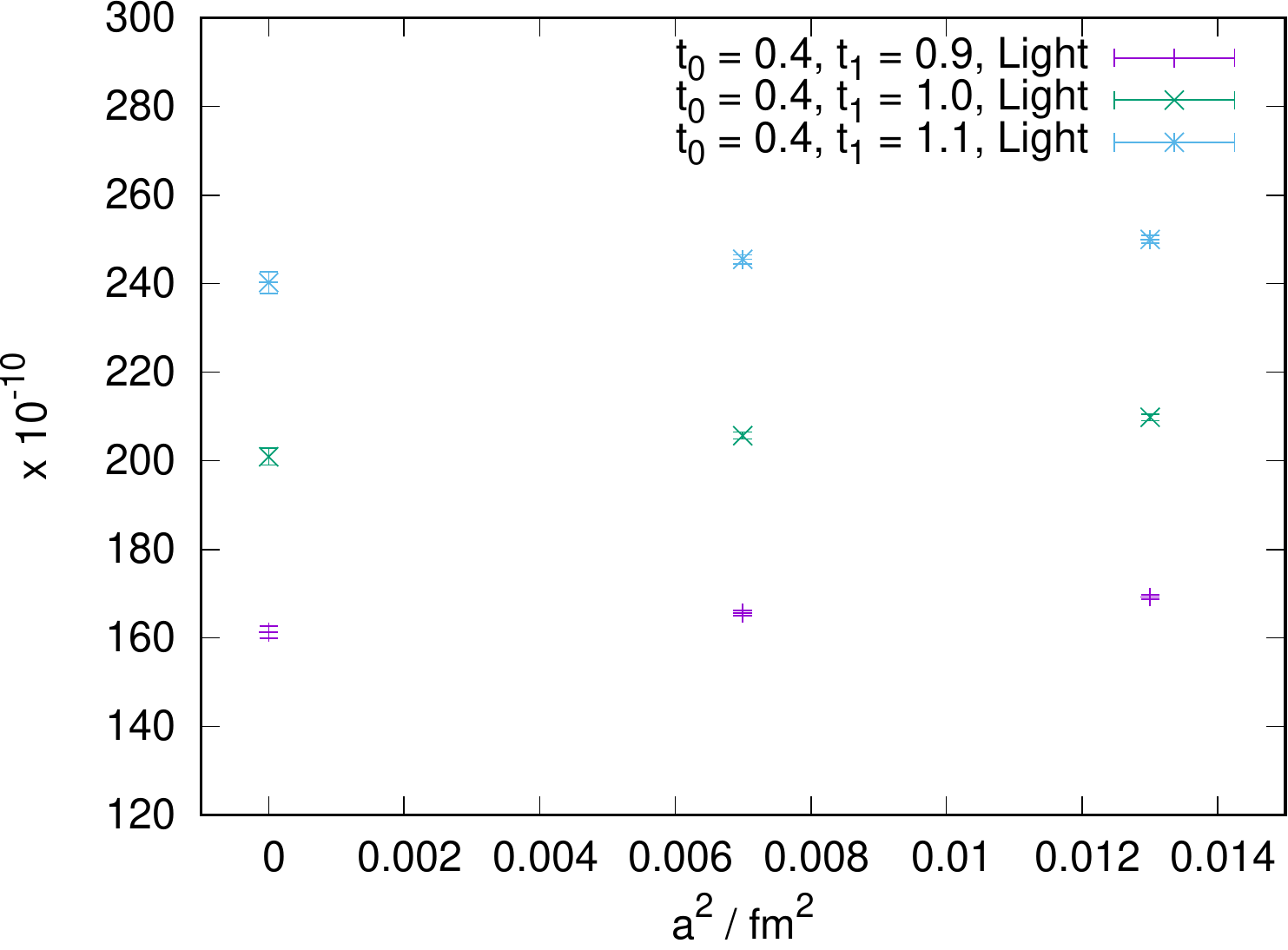}
  \end{center}
  \caption{Continuum limit of light-quark $a_\mu^{\rm W}$ with $t_0=0.4$ fm and $\Delta=0.15$ fm.}
  \label{fig-m9d}
\end{figure}

\skp
{\bf Continuum limit:}
The continuum limit of a selection of light-quark window contributions $a_\mu^W$
is shown in Fig.~\ref{fig-m9d}.
 We note that the results on the
coarse lattice differ from the continuum limit only at the level of a
few per-cent.  We attribute this mild
continuum limit to the favorable properties of the domain-wall
discretization used in this work.  This is in contrast to a rather
steep continuum extrapolation that occurs using staggered quarks as
seen, e.g., in Ref.~\cite{Chakraborty:2016mwy}.

The mild continuum limit for light quark contributions is
consistent with a naive power-counting estimate of $(a\Lambda)^2 =
0.05$ with $\Lambda=400$ MeV and suggests that remaining
discretization errors may be small.  Since we find such a mild
behavior not just for a single quantity but for all studied values of
$a_\mu^W$ with $t_0$ ranging from $0.3$ fm to $0.5$ fm and $t_1$
ranging from $0.3$ fm to $2.6$ fm, we suggest that it is rather
unlikely that the mild behavior is result of an accidental
cancellation of higher-order terms in an expansion in $a^2$.  This
lends support to our quoted discretization error based on an ${\cal
  O}(a^4)$ estimate.  In future work, this will be subject to further
scrutiny by adding a data-point at an additional lattice spacing.

\begin{figure}[t] 
    \begin{center}

      \includegraphics[width=6.9cm]{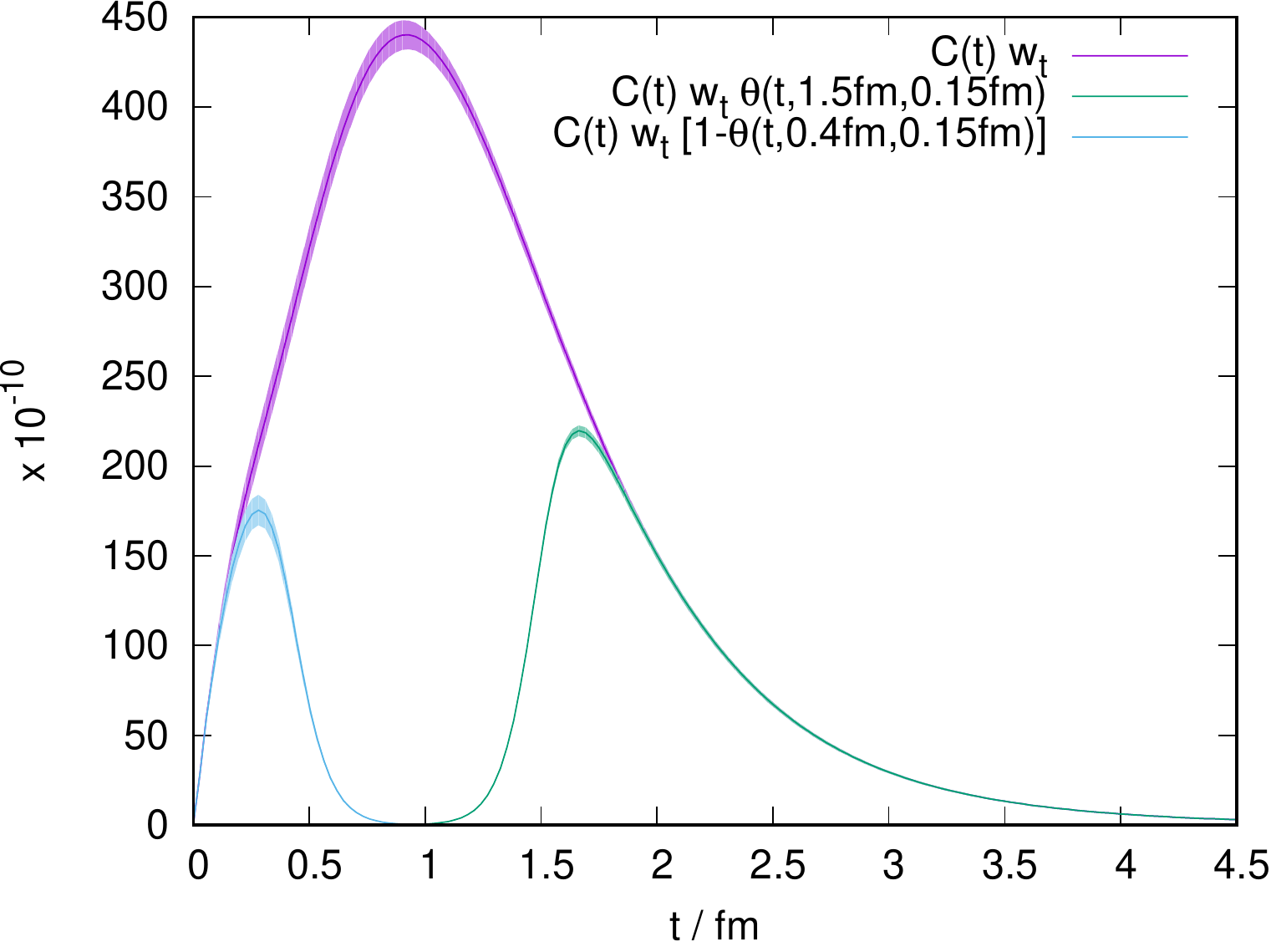}

      \vspace{0.05cm}

      \includegraphics[width=6.9cm]{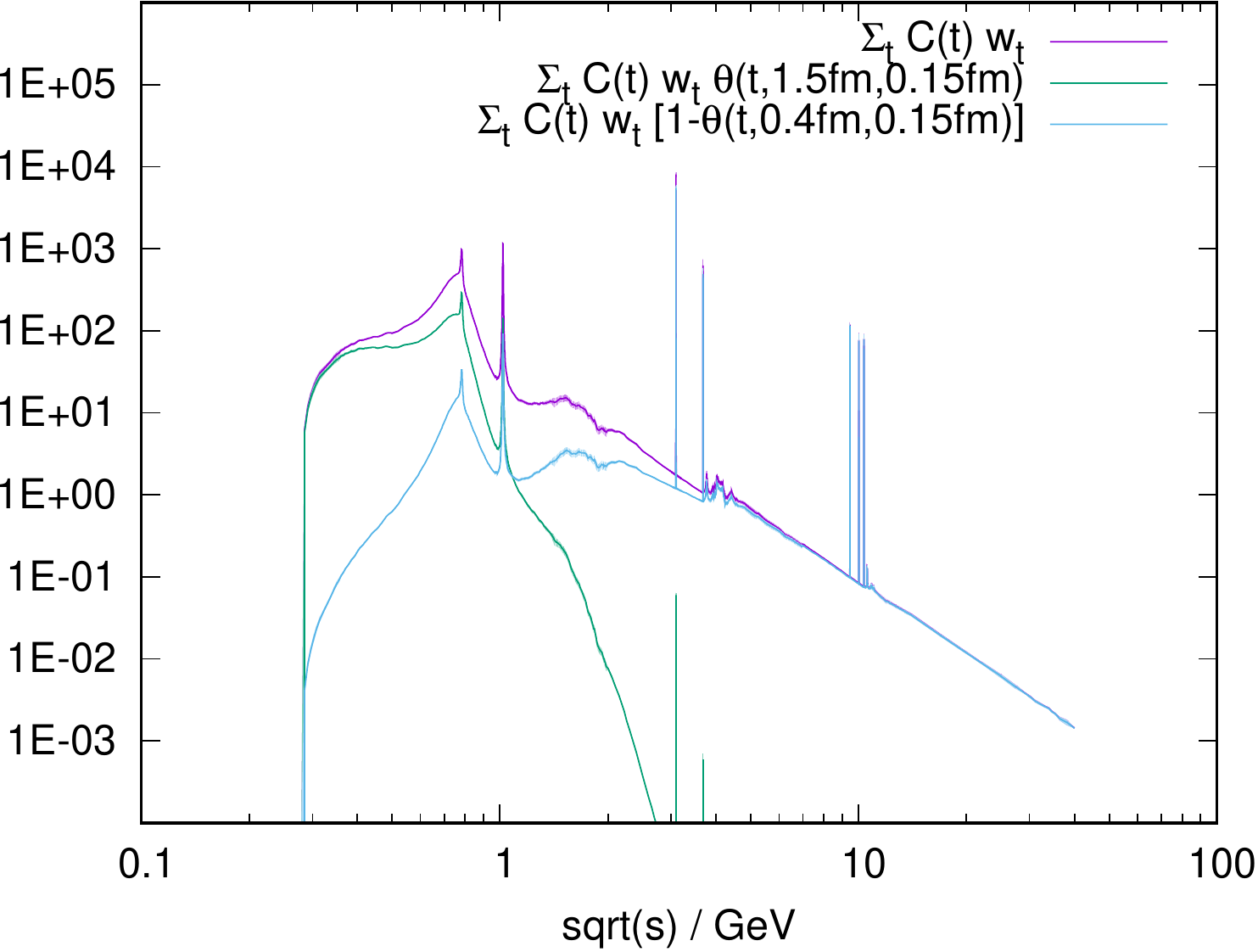}
    \end{center}
  \caption{Window of R-ratio data in Euclidean position space (top) and the effect of the window in terms of re-weighting energy regions (bottom).}
  \label{fig-m9a}
\end{figure}

\skp
{\bf Energy re-weighting:} 
The top panel of Fig.~\ref{fig-m9a} shows the weighted
correlator $w_t C(t)$ for the full $a_\mu$ as well as short-distance
and long-distance projections $a^{\rm SD}_\mu$ and $a^{\rm LD}_\mu$
for $t_0=0.4$ fm and $t_1=1.5$ fm.  The bottom panel of
Fig.~\ref{fig-m9a} shows the corresponding contributions to $a_\mu$
separated by energy scale $\sqrt{s}$.  We notice that, as expected,
$a^{\rm SD}_\mu$ has reduced contributions from low-energy scales and
$a^{\rm LD}_\mu$ has reduced contributions from high-energy scales.
In the limit of projection to sufficiently long distances, we may
attempt to contrast the R-ratio data directly with an exclusive study
of the low-lying $\pi\pi$ states in the lattice calculation.  This is
left to future work.

\skp
{\bf Statistics of light-quark contribution:}
We use an improved statistical estimator including a full low-mode
average for the light-quark connected contribution in the isospin
symmetric limit as discussed in the main text.  For this estimator, we
find that we are able to saturate the statistical fluctuations to the
gauge noise for 50 point sources per configuration.  For the 48I
ensemble we measure on 127 gauge configurations and for the 64I
ensemble we measure on 160 gauge configurations.  Our result is
therefore obtained from a total of approximately $14$k domain-wall
fermion propagator calculations.

\skp {\bf\boldmath Results for other values of $t_0$ and $t_1$:} In
Tabs.~\ref{tab:addtab0}-\ref{tab:addtab6} we provide results for
different choices of window parameters $t_0$ and $t_1$.  We believe
that this additional data may facilitate cross-checks between
different lattice collaborations in particular also with regard to the
up and down quark connected contribution in the isospin limit.

\begin{figure*}[bt]
\begin{center}
  \begin{minipage}{0.45\linewidth}
    \includegraphics[scale=0.55]{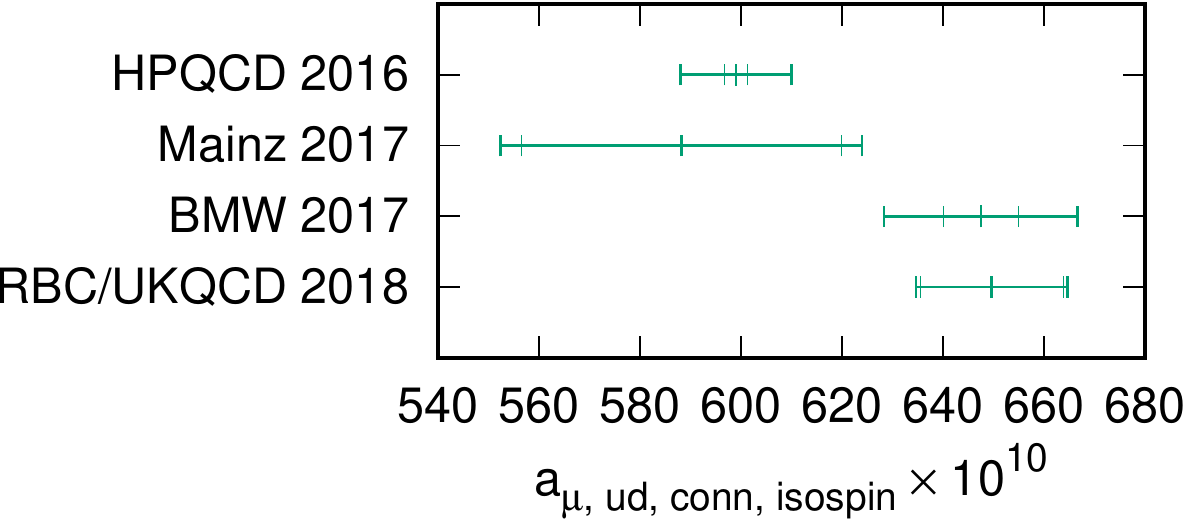}
\includegraphics[scale=0.55]{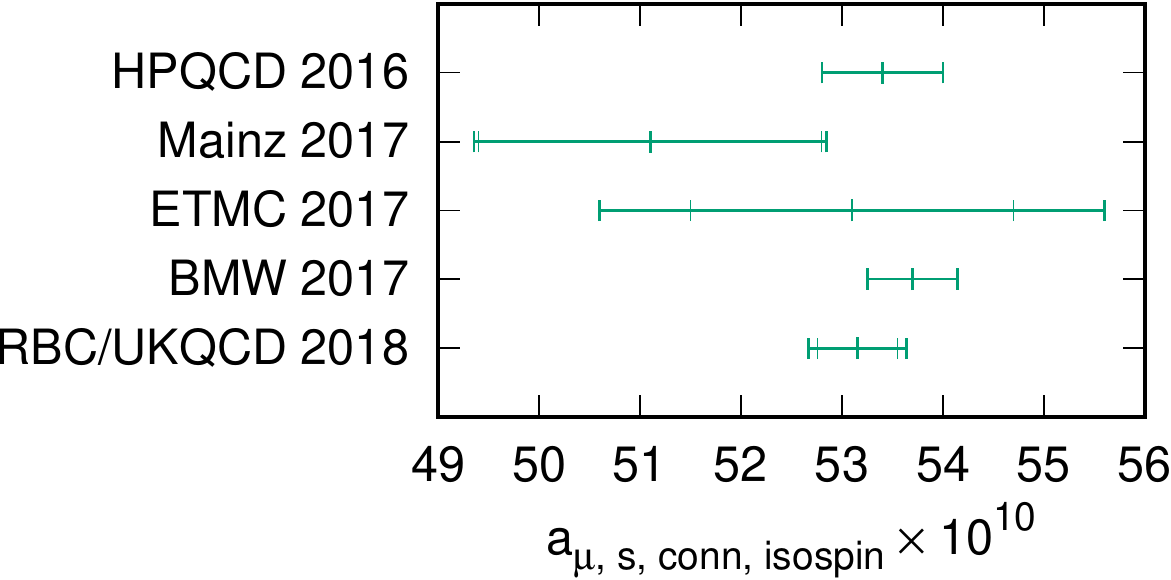}
\includegraphics[scale=0.55]{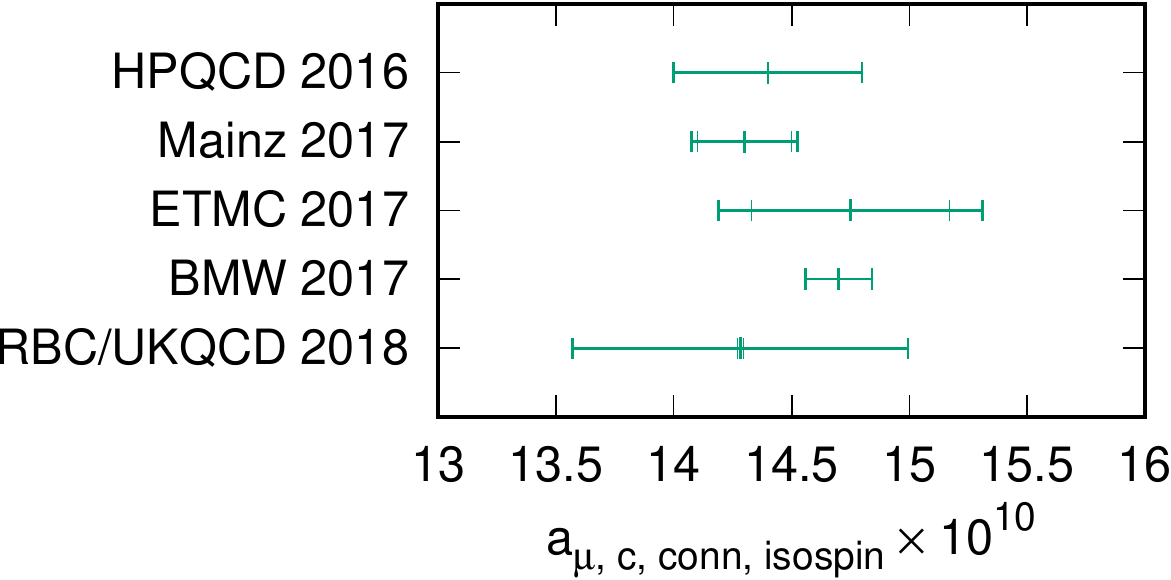}
  \end{minipage}
  \begin{minipage}{0.45\linewidth}
\raggedleft
\includegraphics[scale=0.55]{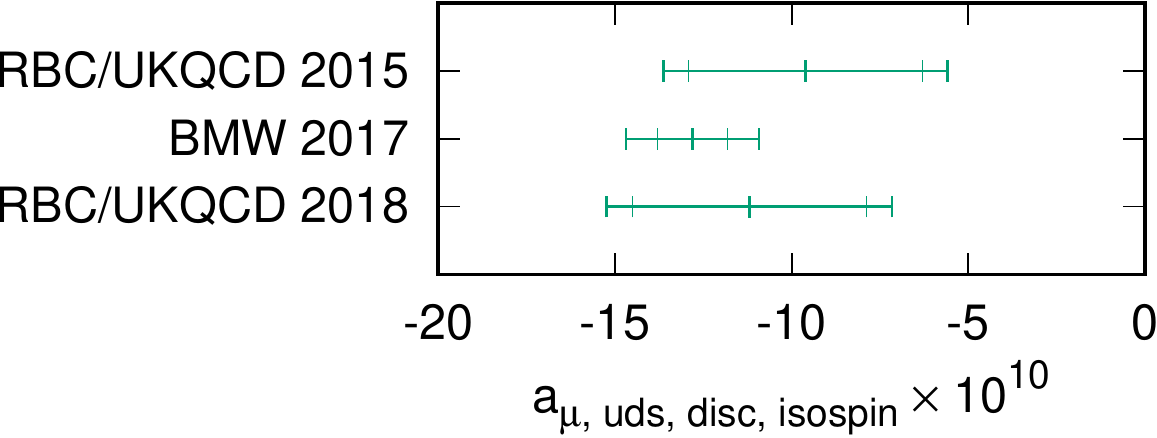}
\includegraphics[scale=0.55]{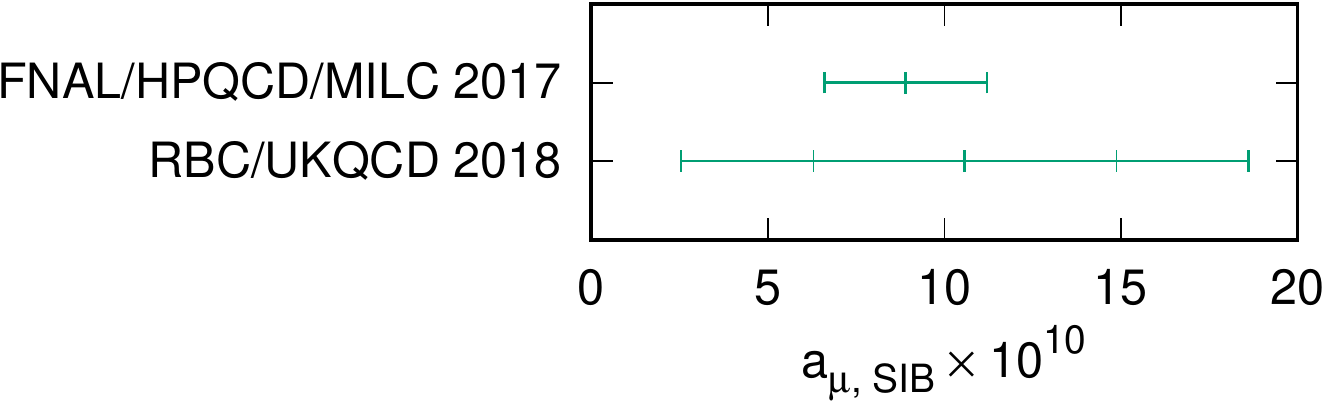}
\includegraphics[scale=0.55]{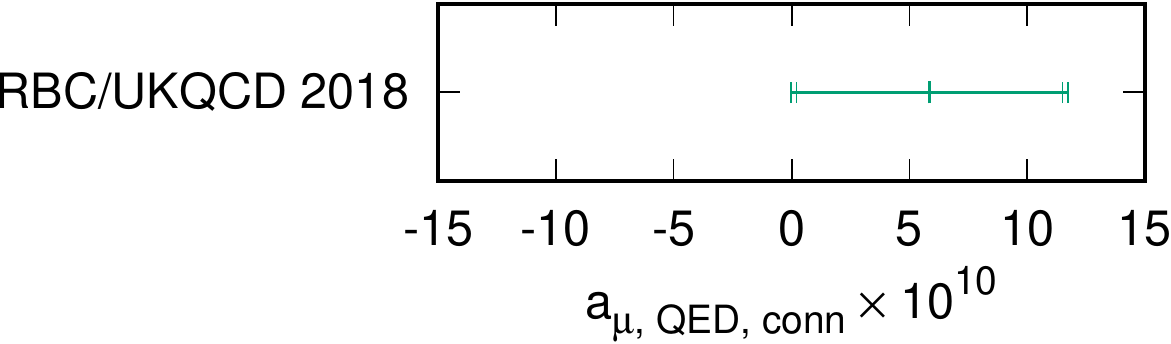}
\includegraphics[scale=0.55]{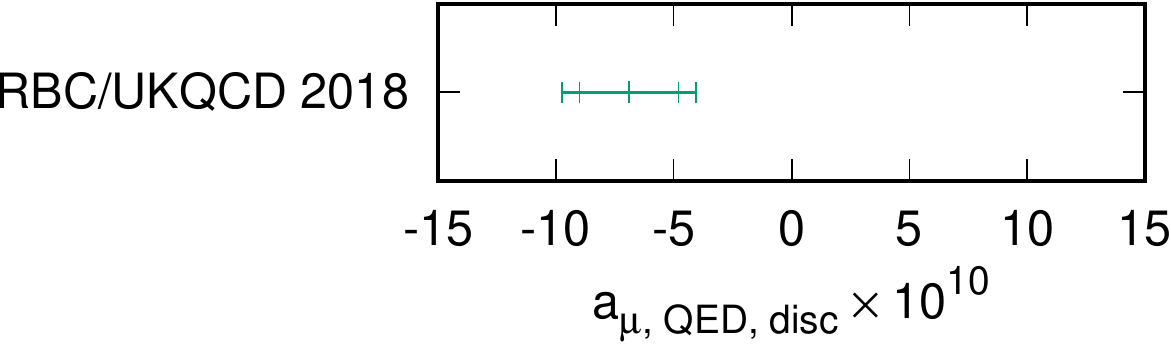}
  \end{minipage}
\end{center} 
\caption{A comparison of our results with previously published
  results.  The references in order of appearance are HPQCD 2016
  \cite{Chakraborty:2016mwy}, Mainz 2017 \cite{DellaMorte:2017dyu},
  BMW 2017 \cite{Borsanyi:2017zdw}, ETMC 2017 \cite{Giusti:2017jof},
  RBC/UKQCD 2015 \cite{Blum:2015you}, and FNAL/HPQCD/MILC 2017
  \cite{Chakraborty:2017tqp}. The innermost error-bar corresponds to
  the statistical uncertainty.}
\label{fig:litcomp}
\end{figure*}

\skp
\skp 
{\bf Comparison of individual contributions:} In
Fig.~\ref{fig:litcomp}, we compare our results for individual
contributions to $a^{\rm HVP~LO}_\mu$ obtained from a pure lattice
QCD+QED calculation to previously published results.  We find good
agreement between the different lattice computations for all results
apart from the up and down quark connected contribution in the isospin
limit.  Further scrutiny of the tension between the HPQCD~2016 and the
BMW~2017 and our RBC/UKQCD~2018 results is desired and will be part of
future work.  As an additional check we have computed the
  small QED correction to the strange quark-connected contribution.
  We find $a_\mu^{\rm s,~QED,~conn}=-0.0149(9)_{\rm S}(8)_{\rm
    C}(30)_{\rm V}\times 10^{-10}$ with error estimates described in
  the main text.  Our result agrees well with $a_\mu^{\rm
    s,~QED,~conn} = -0.018(11) \times 10^{-10}$ of
  Ref.~\cite{Giusti:2017jof}.

\skp
{\bf Bounding method:}
As discussed in the main text, we use a bounding method \cite{LehnerTalkLGT16} 
for the light-quark connected contribution in the isospin symmetric limit.
In the following we give more details for our method and contrast it with the
similar method used in Ref.~\cite{Borsanyi:2016lpl}.  Both our method and the method
of Ref.~\cite{Borsanyi:2016lpl} build on ideas of Ref.~\cite{Francis:2014hoa}.

The correlator $C(t)$ can be written as
\begin{align}
 C(t) = \sum_{n=0}^N c_n e^{-E_n t} 
\end{align}
with real positive energy levels $E_n$ and the constraint that all $c_n \geq 0$.  The
correlator
\begin{align}
  \tilde{C}(t; T, \tilde{E}) =
  \begin{cases}
    C(t) & t < T \,, \\
    C(T) e^{-(t-T) \tilde{E} } & t \geq T 
  \end{cases}
\end{align}
then defines a strict upper or lower bound of $C(t)$ for each $t$
for an appropriate choice of $\tilde{E}$.  For the upper bound,
we proceed as Ref.~\cite{Borsanyi:2016lpl} and use the finite-volume
ground-state energy $E_0$ to define
\begin{align}
 C_{\rm upper}(t) = \tilde{C}(t; T, E_0) \,.
\end{align}
For the lower bound, we use the logarithmic effective mass
\begin{align}
  E^*_T = \log(C(T) / C(T+1))
\end{align}
and define
\begin{align}
  C_{\rm lower}(t) = \tilde{C}(t; T, E^*_T)
\end{align}
in contrast to the choice $\tilde{E} \to \infty$ of Ref.~\cite{Borsanyi:2016lpl}.
It is straightforward to show that
\begin{align}
  C_{\rm lower}(t) \leq C(t) \leq C_{\rm upper}(t)
\end{align}
for all $t$.  This bound is more restrictive compared to the choice of
$\tilde{E} \to \infty$.  Since the effective mass
$E^*_T$ may become noisy at long distances, we also note that
any choice of energy $\tilde{E}$ with $\tilde{E} \geq E^*_T$ provides
a strict lower bound.

\skp
\begin{figure}[bt]
  \begin{center}
    \includegraphics[scale=0.55]{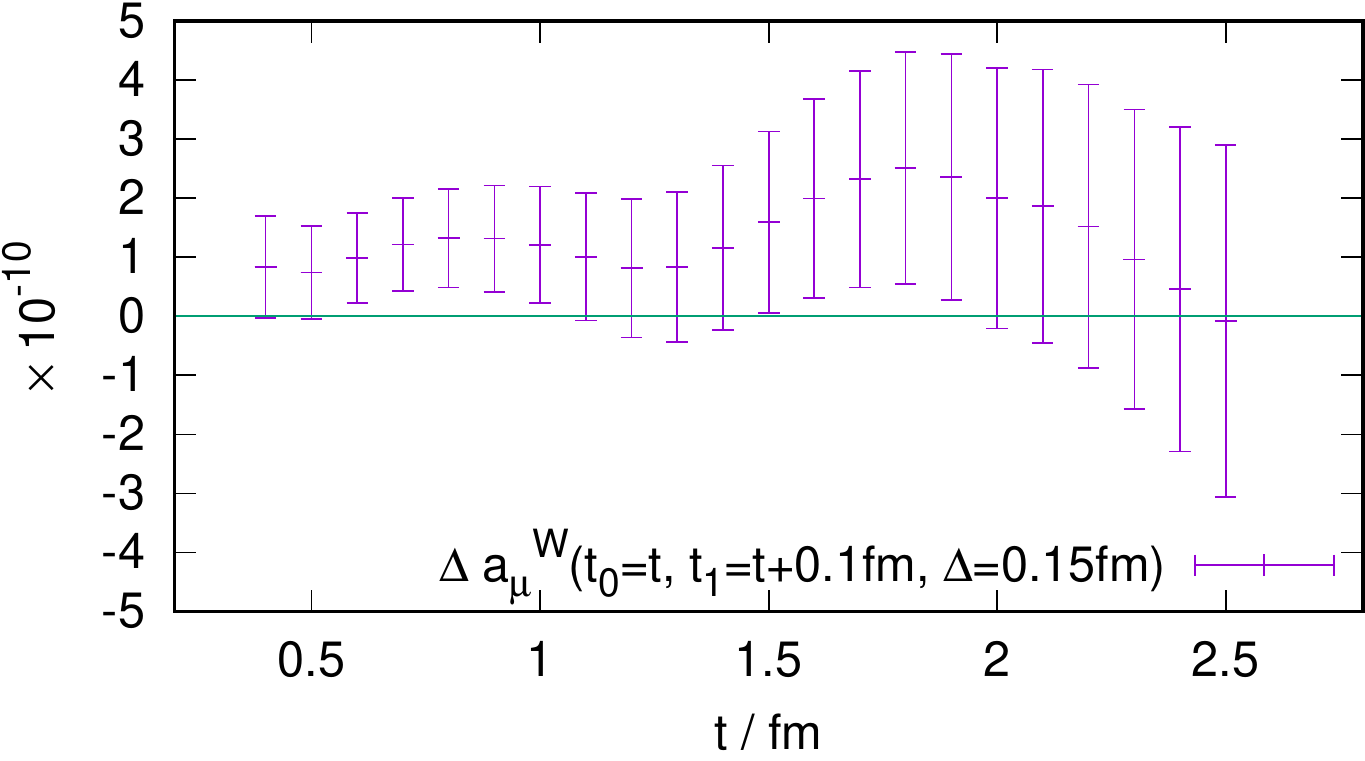}
  \end{center} 
  \caption{The difference of window contributions from the lattice and the R-ratio.
    We show $\Delta a_\mu^W = a_\mu^{W~\rm Lattice} - a_\mu^{W~\rm R-ratio}$.}
  \label{fig:complatr}
\end{figure}

{\bf Consistency of R-ratio and lattice data:}
In Fig.~\ref{fig:complatr} we show the difference of window contributions $a_\mu^W(t_0,t_1,\Delta)$ from the
lattice and the R-ratio with $t_0=t$, $t_1=t+0.1$ fm, and smearing parameter $\Delta=0.15$ fm.
These localized windows are well-defined in the lattice and the R-ratio calculation and allow
for a more precise check of consistency at fixed Euclidean time.  While we find the lattice
calculation to prefer a slightly larger value compared to the R-ratio data of Ref.~\cite{Jegerlehner2017},
this difference is statistically not significant. We will reduce the lattice uncertainties in
the near future in order to provide a more stringent cross-check between both methods.

As noted in the main text, our result for a combined lattice and R-ratio analysis shown in Fig.~7
is based on the R-ratio compilation used in ``Jegerlehner 2017'' but is in better
agreement with the ``HLMNT 2011'', ``DHMZ 2012'', and ``DHMZ 2017'' results
than the pure ``Jegerlehner 2017'' result.  Our value has replaced over one third of
the R-ratio contribution with lattice data and receives its uncertainty in approximate equal 
parts from lattice and R-ratio data.  We are keen on incorporating alternate
compilations of data in future studies and to explore the degree to which the lattice
analysis can help to understand and reduce tensions between the different compilations.

\skp {\bf Estimating QED finite-volume errors:} We estimate the
finite-volume uncertainty of the hadronic vacuum polarization QED
corrections by performing the calculation using an infinite-volume
photon (QED$_\infty$) in addition to the QED$_{\rm L}$ prescription.
We take the difference of both computations as systematic uncertainty
due to the finite volume.  The procedure for both calculations only
differs in the photon propagator that is used.  The QED$_{\rm L}$
prescription uses the photon propagator
\begin{align}
  G_{\rm L}(x) = \frac{1}{V} \sum^\prime_{k} \frac{1}{\hat{k}^2} e^{i k x} \,,
\end{align}
where $\hat{k}^2 = \sum_\mu 4\sin^2(k_\mu/2)$ and $V=\prod_\mu L_\mu$
with lattice dimensions $L_\mu$.  The sum is over all momenta with
components $k_\mu = 2\pi n_\mu / L_\mu$ with $n_\mu \in
[0,\ldots,L_\mu - 1]$ and the restriction that $k_0^2 + k_1^2 + k_2^2
\neq 0$.  For QED$_\infty$ we use instead
\begin{align}
  G_{\infty}(x) = \int_{-\pi}^\pi \frac{d^4k}{(2\pi)^4} \frac{1}{\hat{k}^2} e^{i k x}
\end{align}
with the constraint
\begin{align}
  -\frac{L_\mu}{2} \leq x_\mu < \frac{L_\mu}{2}
\end{align}
for $\mu \in [0,1,2,3]$.

\end{document}